\documentclass[a4paper,twocolumn]{article}
\usepackage[top=2.5cm, bottom=2.5cm, left=2cm, right=2cm]{geometry}
\usepackage{natbib}
\usepackage{aas_macros}
\usepackage{amsmath}
\usepackage{amssymb}
\usepackage{url}
\usepackage{multirow}

\usepackage{graphicx}

\newcommand\ion[2]{#1{\scshape{#2}}}
\newcommand{\samplesize}{2116 } 

\title{Revealing a strongly reddened, faint active galactic nucleus population by stacking deep co-added images}

\author{J\'ozsef Varga$^{1}$\footnotemark, Istv\'an Csabai$^{1,2}$ and L\'aszl\'o Dobos$^{1}$}

\begin{document}



\maketitle

\label{firstpage}

\begin{abstract}

More than half of the sources identified by recent radio sky surveys have not been detected
by wide-field optical surveys. We present a study based on our \textit{co-added image stacking}
technique, in which our aim is to detect the optical emission from unresolved, isolated radio
sources of the Very Large Array (VLA) Faint Images of the Radio Sky at Twenty-cm (FIRST) survey
that have no identified optical counterparts in the Sloan Digital Sky Survey (SDSS) Stripe 82
co-added data set. From the FIRST catalogue, \samplesize such radio point sources were selected,
and cut-out images, centred on the FIRST coordinates, were generated from the Stripe 82 images.
The already co-added cut-outs were stacked once again to obtain images of high signal-to-noise
ratio, in the hope that optical emission from the radio sources would become detectable. Multiple
stacks were generated, based on the radio luminosity of the point sources. The resulting stacked
images show central peaks similar to point sources. The peaks have very red colours with steep
optical spectral energy distributions. We have found that the optical spectral index $ \alpha_\nu $
falls in the range $ -2.9 \leq \alpha_\nu \leq -2.2 $ ($S_\nu \propto \nu^{\alpha_\nu}$), depending
only weakly on the radio flux. The total integration times of the stacks are between $ 270 $ and
$ 300 $ h, and the corresponding $5\sigma$ detection limit is estimated to be about $ m_r \simeq 26.6$ mag. We
argue that the detected light is mainly from the central regions of dust-reddened Type 1 active
galactic nuclei. Dust-reddened quasars might represent an early phase of quasar evolution, and
thus they can also give us an insight into the formation of massive galaxies. The data used in
the paper are available on-line at \url{http://www.vo.elte.hu/doublestacking}. 

\end{abstract}

\renewcommand{\thefootnote}{\fnsymbol{footnote}}
\footnotetext[1]{E-mail: jozsef-varga@caesar.elte.hu}
\renewcommand{\thefootnote}{\arabic{footnote}}
\footnotetext[1]{Department of Physics of Complex Systems, E\"{o}tv\"{o}s Lor\'and University, Pf. 32, H-1518 Budapest, Hungary}
\footnotetext[2]{Department of Physics and Astronomy, Johns Hopkins University, 3400 North Charles Street, Baltimore, MD 21218, USA}

\section{Introduction}

A significant fraction of sources that have been detected in recent
deep radio sky surveys have not yet been detected in the optical
wavelength regime by wide field surveys. From the about $10^6$ objects
identified by the Very Large Array (VLA) Faint Images of the Radio
Sky at Twenty-cm (FIRST) survey \citep{Becker1995,White 1997},
only approximately $30$ per cent have optical
counterparts in the legacy Sloan Digital Sky Survey (SDSS; \citealp{York 2000, Ivezic 2002})
at a limiting magnitude of $ r \sim 22.2 $.
We estimate that for the SDSS Stripe 82 co-added survey, which is
about $ 1.8 $ mag deeper \citep{Abazajian2009}, this detection ratio is
about $42$ per cent. 

The nature of the optically undetected fraction has been the subject
of debate for many years. Recent studies have shown that there
exists a significant population of reddened radio-selected quasars
\citep{Webster 1995,Cutri 2001,Gregg 2002,Richards 2003,White 2003,Glikman 2004,Martinez 2005,Glikman 2007}.
According to the generally accepted unification
model (e.g. \citealp{Antonucci 1993}), a likely explanation for the low
optical luminosity of these quasars is that their central engines are
obscured by an optically thick dust torus.

In this paper, we use a new image stacking technique to investigate
the average optical properties of unresolved faint point sources
from the VLA FIRST survey that have no detected optical counterparts
in the SDSS Stripe 82 (equatorial stripe) co-added data set
\citep{Abazajian2009}.We call our technique \textit{deep co-add stacking}
(DCS) because we stack images that have already been co-added
from multiple observations of the SDSS equatorial stripe. In our
technique, we put strong emphasis on accurate background estimation
and on correcting for selection effects.

\subsection{Image stacking}
\label{sec:stacking}

Co-adding repeated observations and stacking of images of different,
but similar objects can be used to extend the limiting magnitude
and surface brightness limits of surveys. When stacking images of
different extended sources with varying apparent sizes -- depending
on the specific needs -- cut-out images are made, which can be
rotated, resized and/or otherwise scaled together to make the individual
objects overlap. Stacking increases the signal-to-noise ratio
of the images by reducing the photon noise. For undetected sources,
stacking can be successfully used to reveal an average image of extremely
faint sources that otherwise would be beyond the detection
limits of the instruments. The individual objects must be similar to
each other in order to be able to give a physical interpretation to
the average images. The application of image stacking is limited
by a couple of factors, of which the thermal and read-out noise
of the detectors are the most important. However, modern detector
technology makes it possible to stack hundreds of images together,
without significant systematic background noise in the results.

One important disadvantage of image stacking is its limited capability
for statistical analysis. In the optical wavelength regime,
usually only one or a few stacked images per band are obtained
from the individual cut-outs. Bootstrapping techniques can be used
to determine the variance of the measured properties of the stacked
images by sacrificing the surface brightness depth and signal-to-noise
ratio.

When we are interested in faint objects only, it is important to
exclude those pixels of the individual cut-outs from stacking that
are brighter than a carefully determined threshold; otherwise, the
strong signal from the brighter sources or stray light from nearby
bright sources would wash away the very weak signal of the faint
object. Fine-tuned masking techniques are required to address this
problem, as we show in Section~\ref{sec:masking}.

Previously, image stacking has been used successfully to recover
the undetected light from objects that are visible in a given wavelength
range but are too faint in another to detect them directly.
\citet{White 2007} stacked VLA FIRST images of optically identified
quasars, and investigated the radio properties (the correlation
between radio and optical luminosities, radio loudness) of the sample.
\citet{Hodge 2008} identified the faint radio emission from a
diverse sample of quiescent (optically unclassifiable) galaxies from
the SDSS by stacking their FIRST radio images. They showed that
the radio emission from these galaxies can be related to their star
formation rate, or a significant part of these galaxies might harbour
quiescent active galactic nuclei (AGNs) in their cores and their activity
is related to the stellar mass. They continued their analysis
by stacking radio images of luminous red galaxies (LRGs) from
the SDSS. They showed that low radio luminosity AGNs (with
$1400$~MHz flux densities in the {$10 \leq S_{\text{int},1400} \leq 100$ $\mu$Jy} range) are
characteristic of the LRG population, and that an evolution of the
nuclear activity is apparent for redshifts in the range $0.45 < z < 0.6$ \citep{Hodge 2009}.
Radio contribution by star formation was
ruled out by sample selection (i.e. LRGs are passive galaxies with
no significant star formation).

Another use of the stacking method is the detection of faint haloes
of extended objects. \citet{Zibetti 2004} stacked
SDSS optical images of edge-on disc galaxies to detect their stellar
halo components, and they reached a surface brightness level as
low as {$ \mu_r \simeq 31 $ mag arcsec$ ^{-2} $}. Later, they analysed the radial profile
of intracluster light by stacking the SDSS images of $683$ galaxy
clusters. They identified faint emission from intergalactic stars as
far as {$700$ kpc} from the centres of the clusters, by pushing the surface
brightness limit of the stacked images to {$\mu_r \simeq 27.5-30$ mag arcsec$^{-2}$}
in the SDSS r-band \citep{Zibetti 2005}. Also, \citet{Zibetti 2007}
applied their technique to reveal light from \ion{Mg}{ii} absorbers.
\citet{Hathi 2008} used image stacking to determine the average properties
of the surface brightness profiles of very distant, compact galaxies
in the Hubble Deep Field. \citet{Bergvall 2010}
detected the extended red haloes of low surface brightness SDSS
galaxies.Most recently, \citet{Tal 2011} have investigated
the faint stellar haloes of LRGs by stacking SDSS images. They have
found that stellar light in massive elliptical galaxies can be traced
out to about a radius of {$100$ kpc}.

An interesting similarity of these SDSS-based studies is that the
stacked images show extremely red colours. According to \citet{Zibetti 2004}
the $r - i$ colours of the stellar haloes of the galaxies
reach up to about {$0.8$ mag}, which is attributed to very old and/or
metal-abundant stellar populations. As an explanation, for the case
of edge-on disc galaxies, \citet{de Jong 2008} has provided evidence that
the anomalous halo colours are at least partly a result of the underestimated
effect of the extended tails of the point spread function (PSF).

\subsection{Radio--optical cross-identification}

Cross-identification of radio sources with their optical counterparts
is a challenging problem because of the complex morphology of
radio galaxies. Many studies have previously investigated the possibilities
of automatic radio--optical cross-identification and the connections
between the optical and radio properties of extragalactic
objects. \citet{McMahon 2002} automatically identified the optical
counterparts of the FIRST radio sources in the Cambridge Automated
Plate Measurement (APM) scans of the First Palomar Observatory
Sky Survey (POSS-I) plates, and they found about $ 70.000 $
matches. According to them, in the case of the aforementioned
catalogues, positional coincidence is enough to cross-identify radio
point sources with optical sources at false-positive levels lower
than 5 per cent when an association radius of 2 arcsec is used. This
false-positive rate is, of course, only for the particular combination
of catalogues of that study. The rate depends on the astrometrical
accuracy of the catalogues and the density of sources.
\citet{McMahon 2002} also addressed the cross-identification of double radio
sources with their optically observable counterparts. They found
that an optical counterpart is likely to be found halfway between
the components of the radio source pairs.

\subsection{Origins of the unresolved radio emission}
\label{sec:radiopoint}

In Section~\ref{sec:data}, we construct a sample of optically undetected, unresolved
radio sources with integrated $1400$~MHz radio flux above
{$ S_{1400} \geq 1 $ mJy}. We consider several types of radio sources that can
appear unresolved in the VLA FIRST images: galactic pulsars, radio
stars, hotspots of radio jets associated with the outer regions of
radio galaxies and AGNs (including quasars).

A flux density limit of {$1$ mJy} is useful to separate AGNs from low redshift
starburst galaxies. This limit is sufficiently bright to exclude
most of the starburst galaxies, and yet faint enough to permit high redshift
AGNs and classical radio galaxies \citep{Windhorst 1985, Hopkins 2000}.
The 1-mJy limit was used by \citet{Waddington 2001},
who compiled the Leiden--Berkeley Deep Survey (LBDS)
Hercules sample, which is still the largest optically complete sample
of mJy radio sources at $1400$~MHz.

Because we restrict our analysis to the area covered by the SDSS
Stripe 82 ($-50^{\circ} \leq \alpha \leq 59^{\circ}$, $| \delta | \leq 1.26^{\circ}$), the high galactic latitude
and the relatively high selection limit on the radio flux make it very
unlikely that the resulting point sources are pulsars. For example, the
comprehensive Australia Telescope National Facility (ATNF) pulsar
catalogue \citep{Manchester2005} contains only two pulsars within
the footprint of Stripe 82. This number is negligible compared to
the more than $ 20.000 $ unresolved radio objects that are catalogued
in the VLA FIRST data set within the same footprint.

According to \citet{Helfand 1999}, there are 26 known radio
stars brighter than $ 0.7 $~mJy at $1400$~MHz discovered in the FIRST
survey, from which only three sources lie in the Stripe 82 footprint.
Therefore, it is very unlikely that our optically undetected sample
of \samplesize objects contains any radio stars.

Hotspots of radio galaxy jets might also appear as point sources
in FIRST images. They are almost always associated with an easily
detectable nearby elliptical galaxy and they tend to appear in pairs,
on both sides of the galaxies. We have identified many sources
that are very good candidates for such hotspots, but we have excluded
them from the present analysis using a method described in
Section~\ref{sec:first_selection}.

The fourth type of objects that appear as unresolved, isolated
radio sources are the central regions of supermassive black holes
currently accreting matter. In Section~\ref{sec:discussion}, we argue that we detect the
optical light from this type of sources.

\subsection{Reddened radio-loud quasars}

According to the widely accepted AGN unification scheme (for a
review, see \citealp{Antonucci 1993, UrryPadovani1995}, and references
therein), AGNs are divided into two main classes based on their optical
properties. Type~1 AGNs have both broad and narrow emission
lines and show blue optical continua. This implies that the central
regions of these objects are observed directly with no significant
quantity of obscuring material along the light of sight. However,
Type~2 sources show red optical spectra and narrow emission lines
only. It is believed that Type~2 AGNs are viewed from such directions
that the dense dust clouds surrounding the central engines fall
into the line of sight and obscure much of the visible light coming
from the accretion discs. The broad emission lines are associated
with the central regions of the accretion discs, where high velocity
motion is responsible for the broadening of the lines. Polarimetric
studies have revealed that weak, highly polarized broad emission
lines are also detectable in the spectra of Type~2 AGNs. These highly
polarized lines are attributed to the light that originates from the central
regions but is reflected from gas clouds at greater distances from
the centres.

It has been demonstrated previously that there exists a strongly
reddened population of quasars, which are often missed by the conservative
quasar selection criteria of optical sky surveys.
\citet{Ivezic 2002} presented a very comprehensive study about the optical and
radio properties of positionally matched FIRST and SDSS sources.
By analysing the distribution of quasars, they pointed out that the
existence of a significant population of highly obscured quasars
that are visible in FIRST but not in SDSS cannot be ruled out by
the data. \citet{Richards 2003} estimated that about 10 per cent
of the dust-reddened quasars are missing from the SDSS catalogue
because of selection effects. They also estimated that the fraction of
broad absorption line (BAL) quasars rises with redder colour and
can reach 20 per cent for reddened quasars. In the case of BAL
quasars, the BALs are most likely to originate from the fast moving
dust expelled from the central regions by the strong wind of the
nucleus \citep{Hazard 1984, Weymann 1991, Sprayberry 1992, Ogle 1999, Schmidt 1999, Becker 2000, Hall 2002, Trump 2006}.

\citet{Glikman 2007} have confirmed that the selection of quasars
by infrared colours yields a significantly redder quasar population
than selection based on optical surveys. They have also reported,
based on two-frequency radio observations, that the red colours of
these quasars are more likely to be a result of extinction by dust
rather than enhanced synchrotron emission. They have found that
the intrinsic extinction values of these reddened quasars can reach
as high as $ E(B-V) \simeq 2.5 $~mag.

Certain models suggest that AGNs are turned on by galaxy mergers
and that they spend a significant time obscured by the dust
that originates from the merging host galaxies \citep{Sanders 1988,Hopkins 2005}.
In this phase, the AGNs have high intrinsic
luminosities, but they appear as heavily reddened Type 1s (or ultraluminous
infrared galaxies) because of strong obscuration by the
dust of the hosts. The unobscured quasars become visible once the
radiation from the central engines wipes out the obscuring material.

\citet{Urrutia 2009} have investigated the properties of heavily
reddened quasars of a sample compiled from FIRST, the Two-
Micron All-Sky Survey (2MASS; \citealp{Skrutskie 2006}) and SDSS.
The majority of their sample was spectroscopically confirmed, and
they found that a quasar population with high intrinsic extinction
does exist far from the traditional photometric quasar selection
criteria. \citet{Urrutia 2009} measured the reddening values to be
in the range {$ 0.1 < E(B-V) < 1 $ mag}, where the higher values
are more consistent with obscuration by the host galaxies rather
than being a result of the dust torus surrounding the central engines.
They have also found that these quasars are likely to show BALs
of elements at low ionization states. This observation suggests that
these AGNs are at an early period of their active cycles, when strong
winds driving the dust out from the central regions cause the BALs.

For a comprehensive review of the spectral energy distributions
(SEDs) of Type~1 quasars, see \citet{Richards 2006}.

\subsection{Structure of the paper}

In Section~\ref{sec:data}, we describe the radio object sample selection method,
and the source of the optical imaging data. We give details of the
image stacking algorithm in Section~\ref{sec:algorithm}, and we present the results
of the analysis -- including photometry, SED and radial surface
brightness profiles of the stacked objects -- in Section~\ref{sec:results}. Finally,
in Section~\ref{sec:discussion}, we discuss the optical and infrared properties of our
sample.

We use the concordance flat $\Lambda$ cold dark matter (CDM) cosmology
throughout this paper, with parameters of $H_0 = 71$ km s$^{-1}$ Mpc$^{-1}$, $\Omega_M = 0.27$ and $\Omega_\Lambda = 0.73$.

We have developed our own software tools, written in IDL, to
generate cut-out images and masks, and for the entire image stacking
process. The code is available from the authors on request.


\section{Data}
\label{sec:data}

For the present analysis, we use $1400$~MHz radio data from the VLA FIRST survey \citep{Becker1995} and $ u $, $ g $, $ r $, $ i $ and $ z $-band optical images from the SDSS Stripe 82 co-added data set \citep{Abazajian2009}. The aim of our analysis is to detect the faint optical emission of sources that are individually undetected in the optical, but which have significant emission in the radio. We choose our sample selection criteria according to the requirements detailed in the following two subsections.

\subsection{Sample selection from the FIRST catalogue}
\label{sec:first_selection}

We only select those FIRST objects that are within the SDSS equatorial stripe footprint ($-50^{\circ} \leq \alpha \leq 59^{\circ}$, $| \delta | \leq 1.26^{\circ}$). Objects are required to be compact radio sources, i.e. point-like without any extended radio features. The FIRST catalogue assigns a side-lobe probability $ P_S $ to each object based on oblique decision tree classifiers (refer to the FIRST web-site\footnote{\url{http://sundog.stsci.edu/}} for details). We require that the side-lobe probability must be less than $ P_S \leq 0.1 $. In addition, the integrated {$1400$~MHz} fluxes of the objects must be higher than {$S_{\text{int},1400} = 1$~mJy}. We visually inspect all automatically selected sources, as described in Section~\ref{sec:vis_ins}.

\begin{figure}
\includegraphics{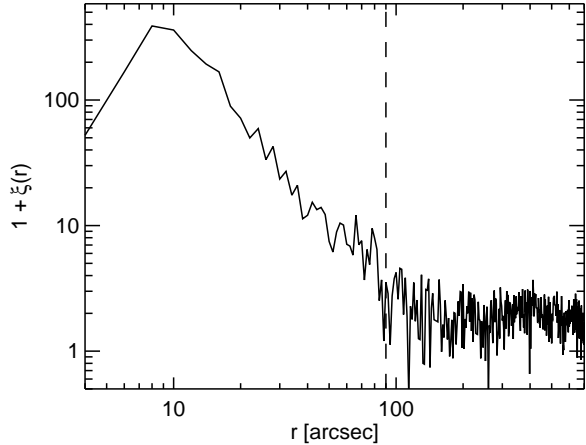}
  \caption{Pair-correlation function of 6019 FIRST radio point sources brighter than {$ S_{\textrm{int},1400} \geq 1 $ mJy}, plotted for small separations only. The reason behind the strong correlation at very small angles is that multiple point sources are likely to be associated with the same radio galaxy. The vertical dashed line indicates our {$1.5$ arcmin} limit on the minimum separation from any other radio sources.}
\label{fig:paircorr}
\end{figure}

In Fig.~\ref{fig:paircorr}, the angular pair-correlation function of the FIRST radio point sources is plotted for small separations. A strong correlation at very small angles is clearly visible, but the function becomes flat above {$1.5$~arcmin}. This behaviour originates from the simple fact that radio galaxies have complex morphologies with many bright features: the multiple lobes and hotspots of radio galaxies are likely to appear as double or multiple associations of unresolved sources in the FIRST images that are close to each other. In most cases, the optical counterparts of these galaxies are easily identifiable in the proximity of the radio sources. In order to make sure that only isolated sources are included in the sample, we consider the angular pair-correlation function of them, and further restrict the selection conditions to exclude radio sources with another close radio companion within \mbox{$1.5$~arcmin}. This criterion is though to effectively exclude most point sources that are associated with optically resolved radio galaxies.

Because we are looking for compact radio sources with no clearly 
identified optical counterparts, in the analysis we only include 
those radio sources that do not have any matches in the SDSS Stripe82
co-added catalogue within {$3$~arcsec}. This condition is much less 
restrictive than the typical astrometric error of SDSS, which is about
{$ 0.1 $~arcsec} \citep{Stoughton 2002}, but it is strict enough to 
exclude any false positive matches. It also excludes extended radio 
sources with small diameters that might show an offset of a few arcsec 
between the optical and radio source positions. This value is chosen 
because practically all true SDSS--FIRST matches have smaller 
separations than {$3$~arcsec} \citep{Ivezic 2002}.

\subsection{Visual inspection of the images}
\label{sec:vis_ins}

After applying the former criteria, 2626 sources remain. We identified these radio sources in the Deep VLA Stripe 82 catalogue \citep{Hodge 2011} which has double the resolution and about three times the signal-to-noise ratio of the FIRST survey. Although the deeper survey does not cover the entire Stripe 82 region (only about 120 square degrees of the total 280 square degrees, 43 per cent), we still can use the deeper catalogue to verify the efficiency of the selection criteria we used to filter the FIRST data set. We are especially interested whether radio point sources can be segregated from extended sources based solely on the information available in the lower resolution FIRST catalogue. Therefore, we visually inspect the FIRST, Deep VLA and SDSS Stripe 82 images of all previously selected sources.

First of all, we exclude all radio objects if the corresponding optical frames are visibly affected by scattered light from nearby bright stars. The number of these objects amounts to about 180. In a few cases, we have to exclude radio objects because they overlap with the outer regions of very bright, extended galaxies.

About 10 per cent of the automatically selected FIRST objects is rejected during visual inspection because they clearly show resolved features or appear very faint and different from the average FIRST PSF.

Next, we compare the FIRST and Deep VLA images of those objects which were observed during both surveys. While all objects appear pointlike in the FIRST images (although some show a slight eccentricity), about 4 per cent of the FIRST point sources are clearly resolved into radio galaxy features in the deeper and higher resolution VLA survey. Unfortunately, there appears to be no way to identify the resolved sources based solely on their FIRST catalogue properties. Hence, exclusion of radio objects that become resolved at higher resolution is only possible in one half of the sample.

\begin{table}
 \centering
  \begin{tabular}{r@{.}l  r@{.}l  r@{.}l r@{.}l}
  \hline
   \multicolumn{2}{c}{R. A. J2000} & \multicolumn{2}{c}{Dec. J2000} & \multicolumn{2}{c}{$S_{\text{int},1400}$} & \multicolumn{2}{c}{rms flux} \\
   \multicolumn{2}{c}{[deg]} & \multicolumn{2}{c}{[deg]} & \multicolumn{2}{c}{[mJy]} & \multicolumn{2}{c}{[mJy/beam]} \\
 \hline
 \hline
331&2879944 & 1&2256938 &  ~~~11&74 &  ~~~~~0&150\\ 
353&8881836 & 1&2255382 &  2&80 & 0&143\\
327&1674805 & 1&2590889 &  5&52 & 0&148\\
353&0435486 & 1&1830354 & 18&99 & 0&147\\
 37&0628471 & 1&1827827 &  2&82 & 0&138\\
 18&8434620 & 1&1813327 &  1&10 & 0&145\\
356&9854126 & 1&1412108 & 14&33 & 0&136\\
\hline
\end{tabular}
  \caption{Sources of the final DCS sample. (The entire table 
is available in machine-readable form at 
{http://www.vo.elte.hu/doublestacking}. A portion is shown here.)}
\label{tab:sample_lst}
\end{table}

\begin{table}
 \centering
  \begin{tabular}{ c c c c c}
  \hline
   $ S_{\textrm{int},1400} $ & median flux & $ \sigma_S $ & $ N $ & $t_{\text{exp}}$ \\
   $ [\text{mJy}] $ & $ [\text{mJy}] $ & $ [\text{mJy}] $ & & $ [\text{h}] $ \\
 \hline
 \hline
 $ 1 $ -- $ 2 $ & ~1.45 & ~0.28 & ~709 & ~290\\
 $ 2 $ -- $ 4 $ & ~2.71 & ~0.54 & ~653 & ~270\\
 $ >4 $         & ~7.61 & 26.1~~ & ~754 & ~300\\
 \hline
 Total          & ~2.78 & 16.7~~ & 2116 & 860 \\
\hline
\end{tabular}
  \caption{Distribution of the DCS sources by subsamples. $ S_{\textrm{int},1400} $ is the {$ 1400 $~MHz} flux density interval of the bin; $ \sigma_S $ is the variance of the flux density, $ N $ is the number of the sources of each subsample; and $t_{\text{exp}}$ the total effective exposure time of the stacked optical images.}
\label{tab:subsampl}
\end{table}

\begin{figure}
\includegraphics{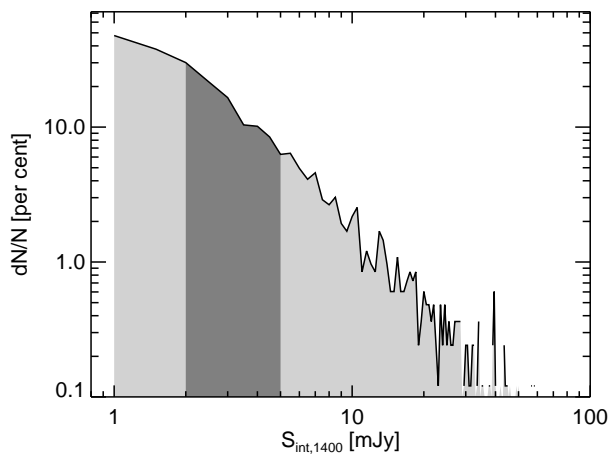}
  \caption{Distribution of the radio flux $S_{\text{int},1400}$ of the DCS sample. The grey areas indicate the three subsamples. light grey on the left: $ 1-2 $ mJy; dark grey: $ 2-4 $ mJy; light grey on the right: $ >4 $ mJy.
  }
\label{flux_hist}
\end{figure}

To select a radio source we required that it should be at least $ 1.5 $~arcmin away from any other FIRST radio sources. When looking at the deeper radio survey images, objects of low radio brightness are apparent within $ 1.5 $~arcmin in a few cases. Also, a very few objects detected in the FIRST survey are missing from the Deep VLA images. We attribute these false detections to glitches in the FIRST observations.

By looking at the higher-quality Deep VLA images, it is estimated
that about $9$ per cent of the FIRST objects would be excluded at the
depth of the Deep VLA observations. However, we have found that
the exclusion or inclusion of these objects does not modify our final
findings significantly; the difference in the measured fluxes is less
than 5 per cent, which is in the range of the estimated error value.
Although it can only be done for half of the data set, we still decide
to exclude them from our final stacks.

\subsection{Subsample definitions}

The final, deep co-add stack (DCS) sample consists of \samplesize objects, listed in Table~\ref{tab:sample_lst}. The sources are grouped into three, roughly equal cardinality subsamples based on their integrated apparent radio flux densities $S_{\text{int},1400}$. The size of the subsamples make it possible to generate stacks with high enough signal-to-noise ratio. The definitions of the subsamples are based on apparent flux because there are no redshift estimates available for the sample. Still, we can expect to see some dependence of the derived parameters on the apparent radio flux at the end of the analysis if the objects lie in a relatively narrow redshift range (i.e. a correlation between apparent and absolute radio luminosities exists).

Table~\ref{tab:subsampl} summarizes the properties of the three DCS subsamples. The distribution of the radio fluxes is plotted in Fig.~\ref{flux_hist}.

\subsection{SDSS co-added images}

For the stacking, we use the SDSS equatorial stripe co-added imaging
data to create cut-outs centred on the selected DCS source
coordinates. The SDSS Stripe 82 images were co-added from about
$20$--$40$ individual exposures \citep{Abazajian2009}.\footnote{The number of 
scans of Stripe 82 now exceeds 70, but only $20$--$40$ scans have been included
in the co-addition (mostly data collected in 2005).} Each exposure
took {$54$~s} and the resulting catalogue has an estimated $r$-band magnitude
limit of about {$ r \lesssim 24 $~mag}, which is about one magnitude
deeper than the rest of SDSS \citep{Gunn 1998}.


\section{Image stacking}
\label{sec:algorithm}

\subsection{Cut-outs}

\begin{figure*}
\includegraphics{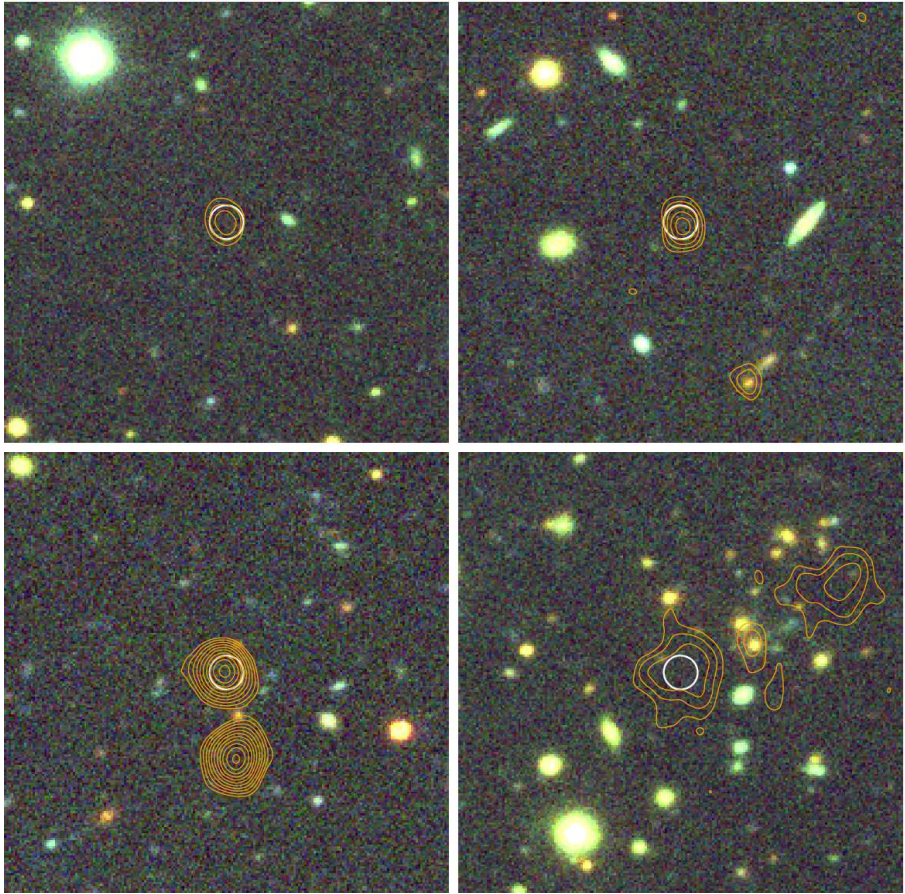}
\centering
  \caption{Sample cut-out images from the SDSS equatorial stripe (Stripe 82) co-added survey with {$ 1400 $~MHz} radio flux contour overlays from FIRST. The top two images are examples of simple, isolated radio point sources, while the lower two ones show radio galaxies with more complex radio features. Only the top left object contributes our sample, the others (including the top right one) are excluded by the criterion which requires that no other radio objects should be in a $1.5$ arcmin radius of a given point source. The images are 80 arcsec wide on the side. The white circles have 3 arcsec radii. The contour levels are $0.4$, $0.65$, $1.0$, $1.45$, $2.0$, $2.65$, $3.4$, $4.25$, $5.2$, $6.25$, $7.4$ mJy beam$ ^{-1} $.}
  
\label{fig:opt_rad_img}
\end{figure*}

The DCS sample consists of radio sources with optical emissions below the detection limit of the SDSS Stripe 82 co-added catalogue. To reveal the average undetected light from these sources, first we generate cut-out images from the already co-added Stripe 82 images centred on the radio coordinates. Cut-outs of the size 200 by 200 pixels are made, which is equivalent to $ 80 $ by $ 80 $ arcsec at the resolution of SDSS. Cut-out image examples with radio flux contour overlays are presented in Fig.~\ref{fig:opt_rad_img}.

Because the background has already been subtracted from the
images during the co-addition, the cut-out images should have sky
levels of 0. Later, we show that this is not the case. In Section~\ref{sec:random},
we introduce a technique to re-estimate the zero level of the cutouts
using an algorithm based on using cut-outs taken at random
coordinates.

\subsection{Masking}
\label{sec:masking}

A significant portion of the pixels of the 200 by 200 Stripe 82 cut-outs is from bright objects that are out of our interest. Also, very frequently, stray light from nearby, bright ({$ m_i < 16 $ mag}) stars contaminates the entire image, but light from interstellar clouds of the Milky Way can also be seen in some of the cut-outs. In order to reveal the very weak signal from the undetected sources, we have to exclude bright pixels efficiently. Our masking algorithm is based on a histogram cutting technique as described below.

\subsubsection{Bright objects}

The $ 1.5 $ arcmin limit on the closest radio source implicitly assures that no associated radio galaxies will appear in the images (c.f. Section~\ref{sec:first_selection}). Also, the condition that the selected radio sources have to be further away than {$ 3 $ arcsec} from any detected optical sources ensures that there will be no readily visible optical sources in the centres of the cut-out images. However, outside the $ 3 $ arcsec radius there can be numerous objects that have to be masked out before stacking the cut-outs together, otherwise the weak signal from the centres would not be detectable.
 
\subsubsection{The masking algorithm}
\label{sec:threshold}

First, we have to determine the pixel values above which pixels
should be rejected. Setting the right value for the masking threshold
is crucial for the stacking to be successful. The values should be
chosen in such a way that all pixels belonging to readily detectable
objects are masked, while sky pixels and pixels of undetectable faint
objects are kept.

\begin{figure*}
\includegraphics{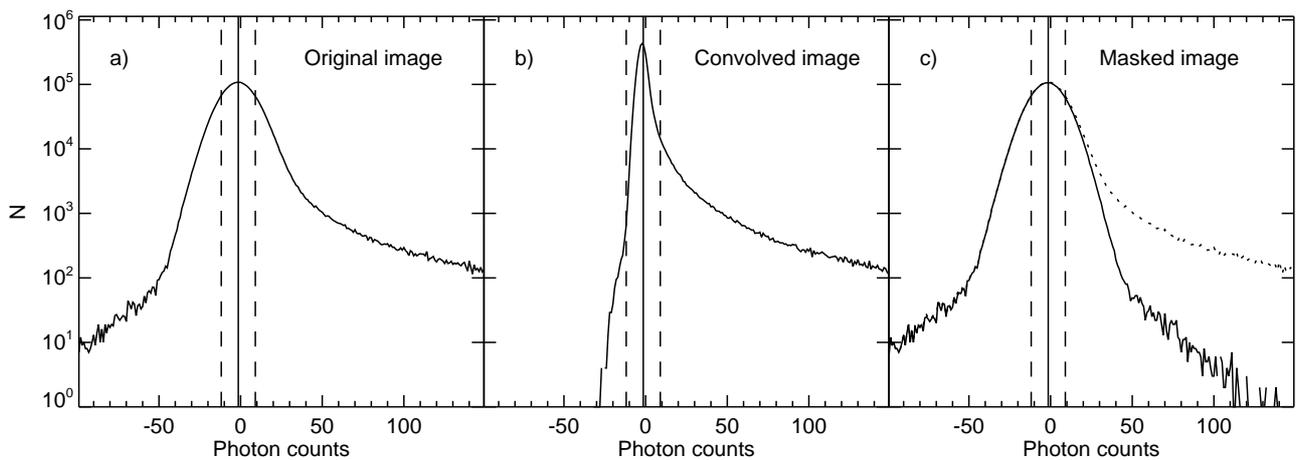}
  \caption{Histograms illustrating the steps of our masking algorithm. Note the logarithmic vertical scale. a) Histogram of the pixel values of a single, typical SDSS Stripe 82 co-added frame. b) Histogram of the same frame convolved with a circular top-hat kernel with a radius of $ 1 $~arcsec. c) Histogram of the frame after applying the mask (solid line). The unmasked histogram is indicated with a dotted line. The solid vertical lines are at the mean, the dashed lines are at the {$\pm 1\sigma$} position of the original distribution. See the text for more details.}
\label{fig:hist}
\end{figure*}

\begin{figure}
\resizebox{8.3cm}{!}{\includegraphics{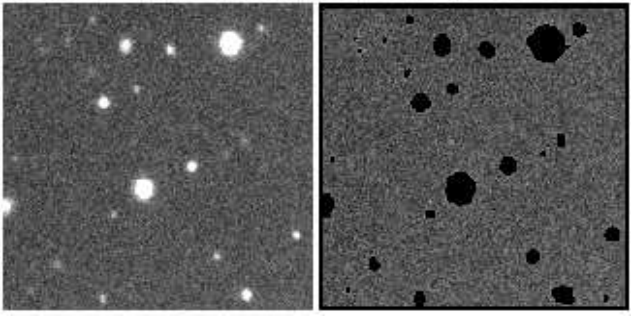}}
  \caption{Left: a sample cut-out image from the stack. Right: The same 
  image overlaid with the mask (shown as black regions).}
\label{fig:mask}
\end{figure}

The histogram of the pixel values of a single Stripe 82 co-added 
frame is plotted in panel (a) of Fig.~\ref{fig:hist} as an example. 
Because we use co-added images, the sky level is already subtracted 
at least to first order. This can be seen clearly from the figure; 
the histogram peaks nearly at zero and background pixels can have 
both positive and negative values. The negative part of the histogram 
is completely dominated by the noise of the sky pixels, while the 
longer tail towards the bright pixels comes from the objects.

The masking process is carried out in four steps, as follows.

\begin{enumerate}
  \item First, we determine the masking threshold. We fit the negative
side of the distribution of pixel values with a Gaussian function,
assuming that the distribution of the background pixels is symmetric
around zero. The masking threshold is then set to be at the $ 1\sigma $
variance of the fitted Gaussian. We indicate the value of the mean
and {$ \pm 1\sigma$} as vertical lines in the panels of Fig.~\ref{fig:hist}.

A higher threshold would let more bright pixels into the statistics,
which would increase the contribution of light from unwanted
objects to the background, resulting in lower contrast stacks. However,
if the value of the threshold was too low, we would filter out
much of the light from the interesting faint sources themselves and
render them undetectable in the stacks. We have found that the $1\sigma$
threshold yields a good balance between contrast and detectability.
In Section~\ref{sec:thresholdrobust}, we show that changing this value slightly does not
affect the final results.

The $\sigma$ values for each SDSS band are determined from the pixel
distribution of the whole ensemble of cut-outs, and the same values
are used throughout the analysis.

  \item Secondly, we make smoothed images from the original cutouts.
This step is important because the imaged sources have diffuse
edges. Therefore, masks derived directly from the cut-out images
by cutting at the masking threshold would not be appropriate. The
masks have to be slightly expanded to cover the smooth edges of
the objects. We also want to avoid the masks being too fragmented.
To achieve these goals, we convolve the cut-out images with a
circular top-hat kernel with a radius of $ 1 $~arcsec. The histogram of a
convolved image is plotted in Fig. ~\ref{fig:hist}(b).

  \item Thirdly, we create the masks from the convolved images;
pixels with values above {$1\sigma$} are masked. Without the initial convolution
of the original images with the top-hat kernel, masked
areas would be strongly non-contiguous because the values of the
masking thresholds are chosen to be very close to the background
noise level. Because the radius of the smoothing kernel limits the
minimum radius of the objects that are masked out, we allow the
kernel to have a radius just slightly larger than the typical seeing
of SDSS. We have found that a kernel radius of less than $ 1 $~arcsec
would make the masks too fragmented. The chosen kernel makes
the masked areas more contiguous and makes them look more or
less circular with sharp edges. A typical mask is presented in Fig.~\ref{fig:mask}.

\item Finally, we generate the masked images by multiplying the
original (non-convolved) cut-outs with the image masks. The histogram
of a masked image is plotted in Fig.~\ref{fig:hist}(c) with a solid line.
The dotted line shows the original, unmasked histogram. It is clear
from the figure that the original symmetric distribution of the background
is preserved during the masking, while object pixels are
effectively masked out.
 
\end{enumerate}

\subsection{Stacking}
\label{sec:stacking_details}

We create the stacked images from the co-added Stripe 82 cut-outs centred on the DCS source coordinates. The cut-outs for each SDSS band are grouped together into three stacks based on the radio flux density of the DCS objects as described in Section~\ref{sec:data}. First, bright pixels are masked out as explained in Section~\ref{sec:masking}, then the remaining pixel values are averaged. Also, the standard deviation of the pixel values at every pixel coordinate is determined to characterize the error.

Stacked images are calculated according to the following formula:
\begin{equation}
\label{eq:stack}
	s_{kl} = \frac{\sum_{i=1}^{N} p_{ikl} w_{ikl}}{\sum_{i=1}^{N} w_{ikl}} - c_{kl}
\end{equation}
Here $ s_{kl} $ is the pixel value of the stack, $ p_{ikl} $ is the pixel value of the $ i^{\text{th}} $ cut-out image and the subscripts $k$ and $l$ index the pixel coordinates ($k, l = \left[ 1,200 \right]$), $ w_{ikl} $ is the mask, its value is 0 if the pixel is masked out and 1 if it is not; $ N $ is the total number of cut-outs. It has been mentioned previously that, although the sky level was subtracted from the Stripe 82 co-added images, the average background levels of the cut-outs usually show a small bias. The correction term $ c_{kl} $  is introduced to correct for this bias. We will explain this last term in detail in Section~\ref{sec:random}. 

\begin{figure*}
\centering
\resizebox{14cm}{!}{\includegraphics{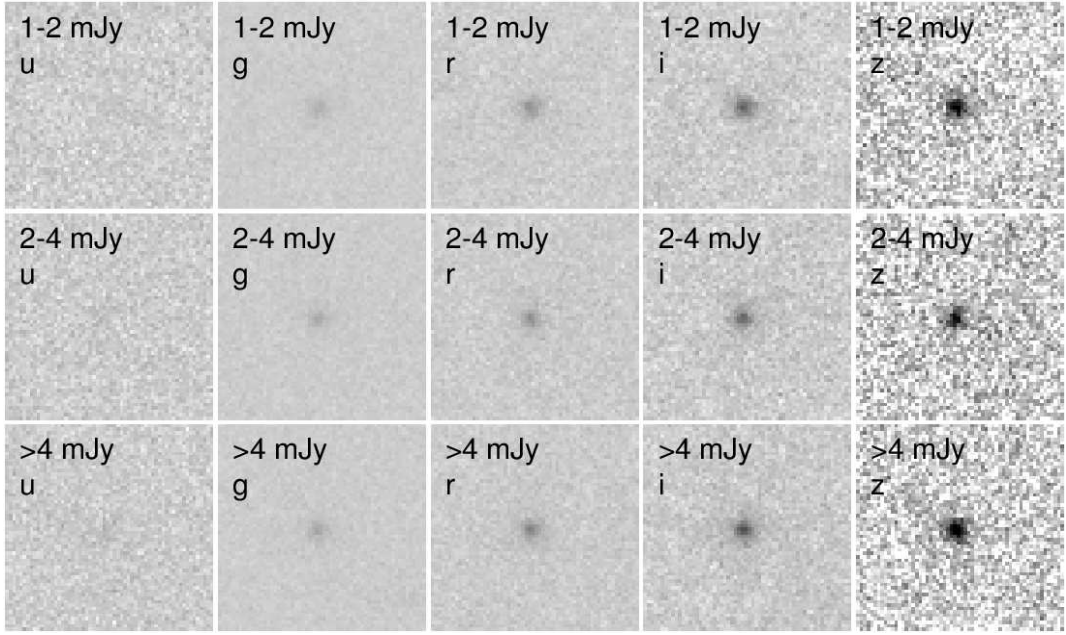}}
  \caption{The stacked images of the radio-selected cut-out images, for 
each radio flux bin and SDSS band. We use the same inverted logarithmic 
scaling for all images. The width of the images is 24 arcsec.}
 \label{fig:stacks}
\end{figure*}

In Fig.~\ref{fig:stacks}, we plot the stacked images for every radio flux bin and SDSS imaging band. A PSF-like central object is clearly visible in the centres of the images. We attribute these peaks to the optical emission from the radio-selected objects.

\subsection{Sky level recalibration}
\label{sec:random}

The SDSS imaging pipeline \citep{Stoughton 2002} estimates the local sky level using clipped median in 256 by 256 pixel boxes, centred on every 128 pixels of the 2048 by 1489 pixel frames\footnote{For more details, see \\
\url{http://www.sdss.org/dr7/algorithms/sky.html}}. A sky constructed by interpolation of these local sky estimates is then subtracted from the frames, prior to co-adding the frames of different scans. The resolution of the sky estimation does not allow for complete elimination of the light from high latitude galactic clouds due to their finer texture. Although the imaging pipeline attempts to subtract the wings of the PSF of saturated stars, the algorithm is not too aggressive. Wings of stars can easily fill the $ 80 $ by $ 80 $~arcsec cut-outs entirely. These two effects can significantly increase the average background level and reduce the contrast of the stacked images, thus need to be eliminated.

The cut-outs we generate show a tiny bias of the background level 
towards negative values. Because we are dealing with objects that 
are extremely faint in the optical bands (in the range of {$ 30 $--$ 300 $~nJy~arcsec$ ^{-2} $} 
or {$ 25 $--$ 28 $~mag~arcsec$ ^{-2} $}), this small bias (estimated to be 
about {$ 10 $--$ 60 $~nJy~arcsec$ ^{-2} $} or {$ 27 $--$ 29 $~mag~arcsec$ ^{-2} $}, depending on the 
photometric filter used) would significantly affect the photometry 
of the stacked images. The bias is likely to be the consequence of a 
remaining gradient in the background level in the whole co-added 
Stripe~82 frames, which becomes apparent if only small parts of 
the frames are considered. Also, the algorithm applied to co-add 
the Stripe 82 images has used $2.326\sigma$ clipping to determine the sky 
level, while we use a much restrictive $1\sigma$ masking threshold. As a 
consequence, the sky level of the cut-outs has to be re-estimated. 
To do this, we simply subtract the average value of the unmasked 
pixels from the cut-outs on a per image basis.

\subsection{Selection bias of the stack sky levels}

When composing the DCS sample, we imposed selection criteria based on the separation of the radio sources from the neighbouring optical sources. This introduces a strange bias to the background of the cut-outs, as we will show in this section.

To investigate the issue, we generate stacked images of cut-outs centred on random coordinates. We call these special stacks \textit{random stacks}. First, for every DCS source coordinate pair, a random coordinate pair is generated within the same Stripe 82 frame where the DCS object is. We require these random points to satisfy the same selection criteria as the DCS objects, i.e. they are at least $ 3 $ arcsec away from any optically detected Stripe 82 co-added sources and at least $1.5$ arcmin away from any other FIRST radio sources. Second, random cut-outs are masked as described in Section~\ref{sec:masking}. Third, the pixel values are averaged according to the following formula:
\begin{equation}
	s^\textrm{R}_{kl} = \frac{\sum_{i=1}^{N} p^\textrm{R}_{ikl} w_{ikl}}{\sum_{i=1}^{N} w_{ikl}},
\end{equation}
where $ s^\textrm{R}_{kl} $ denotes the pixel value at the $ k $ and $ l $ pixel coordinates of the random stack image (the upper R is used to identify that this is a random stack); $ p^\textrm{R}_{ikl} $ is the pixel value of the random cut-outs, where $ i $ indexes the individual cut-outs; $ w_{ikl} $ is the mask, its value is 0 is the pixel is masked out and 1 if it is not; $ N $ is the total number of cut-outs, as before. Separate random stacks are generated for all five SDSS bands and for all three DCS subsamples. (Note, that it is important that the random coordinates are chosen to be within the same Stripe 82 frame as the original DCS source coordinates are, so different random stacks are necessary for each subsample.)

To characterize the variance of the random stacks, we generate 50 of them using different sets of random coordinates. To get the correction term $c_{kl}$ introduced in Section~\ref{sec:stacking_details}, we combine the 50 random stacks into a \textit{super-stack}. The super-stack (denoted as $c_{kl}$ in Eq.~\ref{eq:stack}) is simply the pixel-wise average of the 50 random stacks.

\begin{figure}
\includegraphics{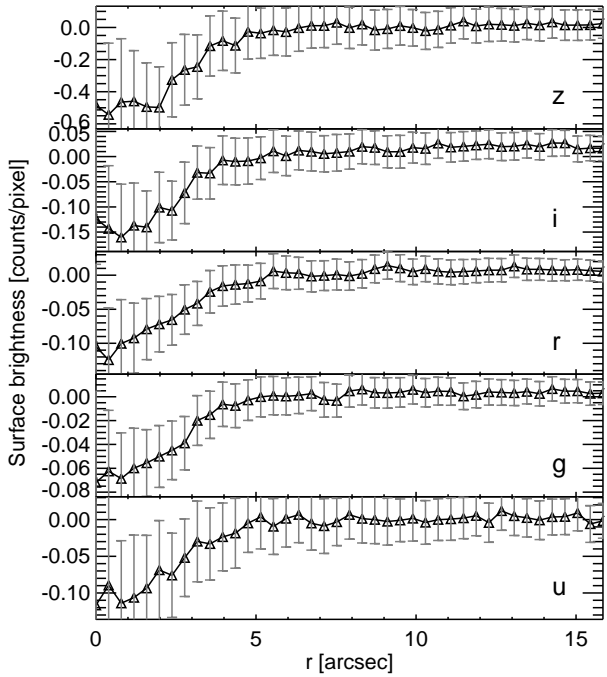}
  \caption{Radial surface brightness profiles of the random super-stacks (the average of 50 random stacks per imaging filter), created for the coordinates of the $ 1 $ -- $ 2 $ mJy subsample using $1\sigma$ masking threshold. The letters of the imaging filters are indicated next to the curves. Error bars show the variance of the values in each annulus.}
\label{fig:randomrad}
\end{figure}

The pixels of these final random super-stacks are once more averaged in annuli centred on the middle of the images to determine the radial surface brightness profile of the background. The resulting profiles are plotted in Fig.~\ref{fig:randomrad}.

The significant feature of the radial surface brightness profiles are the dark spots at the centres of the images. They are very well visible even though variance is higher at these small radii due to the lower pixel counts. We account these dark spots to the selection criterion that restricts coordinate selection to coordinates that are at least $ 3 $~arcsec away from the centres of the closest Stripe 82 objects. The plateaus of the surface brightness profiles of the random stacks come from the faint light of nearby objects. Contribution to these plateaus from stray light from farther bright objects is also expected.

\section{The reliability of the stacking method}
\label{sec:robust}

Simple averaging is known to be sensitive to outliers, thus reliable masking of the bright pixels of cut-outs prior to averaging is inevitable. We apply three tests to confirm the robustness of our masking and stacking method: a) jackknife analysis of the entire stacking process, b) analysis of the dependence of the magnitude of the central peaks on the masking threshold, c) analysis of the dependence of the pixel value histogram of the stacked images on the masking threshold.

\subsection{Jackknife analysis}

To estimate the error introduced by possible outliers, we perform a jackknife analysis of the stacking method by repeating the entire processing numerous times, but leaving out a single cut-out image in every iteration.

\begin{figure}
\includegraphics{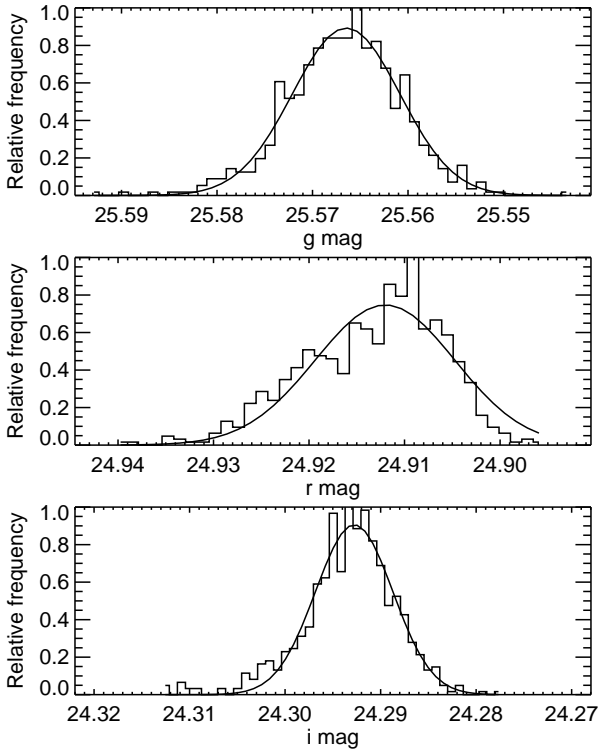}
  \caption{Distribution of the magnitude of the central peak from the 
jackknife stacks (histogram). The means of the fitted Gaussians (solid lines) are 
slightly offset to the right (see text).}
\label{fig:jackknife}
\end{figure}

Fig.~\ref{fig:jackknife} shows the relative frequency of the estimated magnitudes of the central peaks for $ 709 $ jackknife runs for the $g$, $r$ and $i$ band images of the $1$ -- $2$~mJy radio flux bin. The distributions are slightly biased towards the fainter magnitudes as a result of a few significant outliers. The few objects responsible for the faint-end wing of the distribution affect the stacked flux by approx $1$ per cent, thus they have a flux roughly a factor $8$ larger than the average considering a sample of $709$ objects.

The statistical errors of the magnitudes based on the jackknife analysis turned out to be in the $0.1$--$0.2$ mag range. In Section~\ref{sec:photometry}, we will use these error estimates to characterize the uncertainties of the photometry.

\subsection{Effect of the masking threshold}
\label{sec:thresholdrobust}

Masking is required to eliminate the bright pixels of the cut-outs in order to avoid the contamination of the stacked images from moderately bright, but still detectable images -- we are only interested in the undetected, faint objects. In Section~\ref{sec:threshold}, a masking threshold was introduced. The masking threshold is determined from the pixel value distribution of the co-added images by multiplying the variance of the pixel values with a well-chosen number. Here we show that setting the threshold at $1\sigma$ is a good choice.

\begin{figure}
\includegraphics{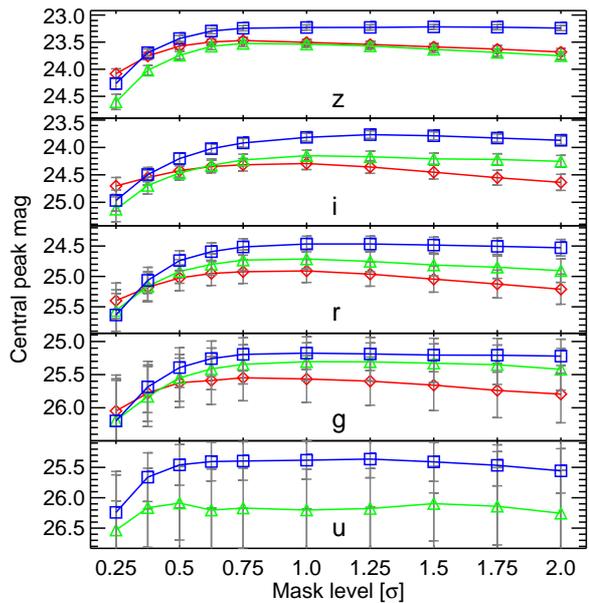}
  \caption{Measured flux of the central peak in the stacked images as a 
function of masking threshold in units of $\sigma$. The panels show the five SDSS 
bands, while colours refer to the radio luminosity bins (red diamond: $1$--$2$~mJy,
green triangle: $2$--$4$~mJy, blue square: $>4$~mJy).}
\label{fig:mask_mag}
\end{figure}

Fig.~\ref{fig:mask_mag} shows the measured magnitudes of the peaks detected in the centres of the stacked images for each SDSS band and radio luminosity bin as a function of the masking threshold. It is observable that the initially increasing curves reach a plateau around $1\sigma$ and then start to decrease slowly. The explanation is simple: If the masking threshold is too low, we lose many pixels belonging the the investigated faint objects. On the other hand, if the threshold is too high the background level of the image is increased by the pixels belonging to other, nearby objects causing the contrast of the stacked images to be lower and the measured magnitudes to be fainter. The maxima of most of these curves are usually located in the $0.75\sigma$ -- $1.25\sigma$ range, so setting the masking threshold at $1\sigma$ is obvious.

Note that if we were to use a smoothing kernel of a different radius
to construct the masks, the start of the plateau of the curves in Fig.~\ref{fig:mask_mag}
would be at different $\sigma$ values. Increasing the radius of the kernel
would move the beginning of the plateaus to the left. This behaviour
is a direct consequence of the masking process; a mask created with
a larger kernel would let more pixels from nearby faint, point-like
objects into the stack at a given masking threshold, because a wide
kernel would smooth out these objects to such an extent that their
pixel values would become less than the masking threshold. This
is why it is important to use a smoothing kernel not much larger
in diameter than the size of the faintest identifiable objects of the
co-added images. This diameter is roughly the FWHM of the PSF.

\subsection{Flux deficit in the central aperture}
Because the masks are uniform in the whole cut-out image area,
some pixels at the central source location can also be masked. This
can introduce a slight flux deficit in the stack because of some
marginally detected sources being masked out; ca. $1$ -- $5$ per cent of
the central pixels are lost, depending on the filter ($u$ is the lowest, $i$
is the highest). If we did not mask in the centre, some contaminating
outlier pixels would easily get into the stack, causing a significant
bias in the photometry (because we would overestimate the flux
of the stacked source). So, this is a necessary trade-off in order to
ensure the photometric quality of the stacked images.

\subsection{Effect of possible outlier pixels}

\begin{figure}
  \includegraphics{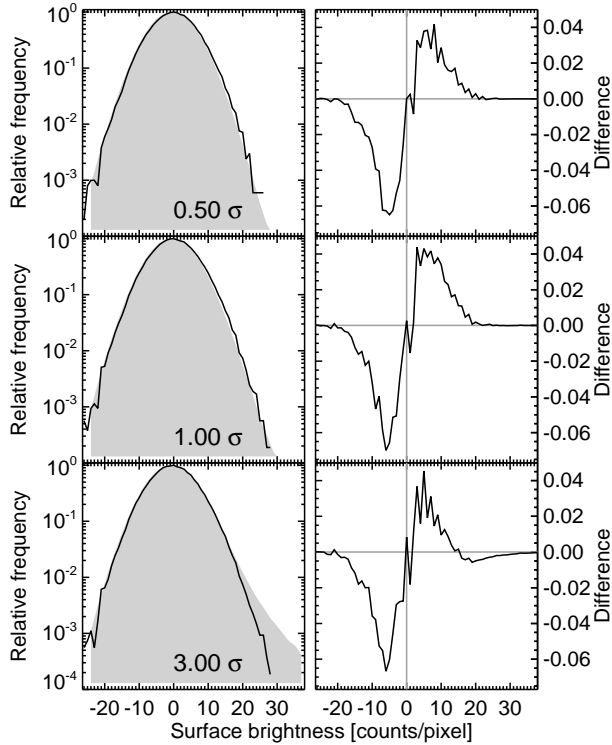}
  \caption{Left column: Distributions of pixel values of the $r$-band cut-outs in the $1$--$2$~mJy radio luminosity bin inside (solid curves) and outside (grey area) of a central circular aperture with a radius of $2$~arcsec for three different values of the masking threshold (indicated in each panel). Panels in the right column show the difference of the two distributions: the distributions of the background pixels subtracted from the distributions of the pixels within the apertures centred on the radio objects. }  
  \label{fig:cutout_histo}
\end{figure}

It is important to show that the signal emerging from the centres of the cut-outs after averaging comes from real objects and not just from a few outlier pixels. Cut-out images after masking will contain sky pixels (background), pixels of nearby faint objects that are above the background level, and pixels near the centres of the cut-outs actually belonging to the radio sources of our interest. The pixel values of these later ones are very close to the background noise level. In Fig.~\ref{fig:cutout_histo} (left column), we plot the distribution of pixel values of the $r$-band cut-outs in the $1$--$2$~mJy radio luminosity bin inside and outside of a central circular aperture with a radius of $2$~arcsec for three different values of the masking threshold. The solid lines (shaded areas) show the normalised distributions of the pixel values inside (outside) the central aperture. The panels in the right column show the difference of the distributions: the distributions of the background pixels subtracted from the distributions of the pixels within the apertures centred on the radio objects. There are two important things to observe in Fig.~\ref{fig:cutout_histo}: a) the distribution of background pixels have a long tail towards bright pixels, and b) the distribution of the pixels inside the aperture is slightly shifted to the right with respect to the background.

The long tail of the distribution of background pixels comes from the relatively bright pixels of nearby faint objects, like the unmasked diffuse envelopes of nearby galaxies. The central aperture does not contain such pixels because of the selection criterion we used to select only those radio sources which are separated from the nearest optically detected objects by at least $3$~arcsec.

The small shift between the peaks of the distributions is due to the actual signal we are looking for. Optical emission from the radio objects is so faint that the individual pixels cannot be distinguished from the noise of the background, but stacking the cut-outs reveals the signal.

The lack of the bright tail of the pixel value distribution of the central pixels and the small shift between the distributions of background pixels and central pixels confirm our claim that the detected peaks in the stacked images come from numerous faint objects and not from just a few brighter (but still faint) outlier objects.

\section{Results}
\label{sec:results}

Altogether 15 stacked images were created, one for each SDSS filter and for each DCS subsample. In Fig.~\ref{fig:stacks} we have already plotted the images of the stacks. The PSF-like point sources are clearly visible at the centres, and we consider them to be the optical signal coming from the FIRST radio sources. The peaks show a monotonically increasing intensity towards the longer wavelength bands, that is the sources are red in the optical colours.

\subsection{Signal-to-noise ratios}

The signal-to-noise ratios and rms background noise levels of the 
stacks are summarized in Table~\ref{tab:snr}. These signal-to-noise ratios are 
sufficient to calculate the photometric properties of the central 
peaks, except for the $u$ filter.

\begin{table}
 \centering
  \begin{tabular}{c r@{.}l  r@{.}l  r@{.}l r@{.}l}
  \hline
          & \multicolumn{6}{c}{S/N ratio} & \multicolumn{2}{c}{rms noise} \\
   Filter & \multicolumn{6}{c}{$S_{\text{int},1400}$ bin [mJy]} & \multicolumn{2}{c}{$\left[\text{mag}~\text{arcsec}^{-2}\right]$} \\ 
	      & \multicolumn{2}{c}{$1-2$} & \multicolumn{2}{c}{$2-4$} & \multicolumn{2}{c}{$>4$} & \multicolumn{2}{c}{ } \\
 \hline
 \hline
 $u$ &  0&6 & 1&1 & 1&7 & ~~~~~~~~29&0\\
 $g$ &  5&7 & 6&1 & 7&4 & 30&0\\
 $r$ &  6&9 & 7&4 & 9&0 & 29&5\\
 $i$ &  8&5 & 8&7 & 10&8 & 29&0\\
 $z$ &  5&9 & 5&5 & 7&4 & 27&6\\
 \hline
\end{tabular}
  \caption{Signal-to-noise ratios and rms background 
noise levels of the stacked images.}
\label{tab:snr}
\end{table}

\subsection{Photometry}
\label{sec:photometry}

The SDSS catalogue uses a special magnitude system designed for
noisy imaging \citep{Lupton 1999}. The definition of the
so-called `luptitudes' is based on the $ asinh $ function instead of the
traditionally used logarithm, and it contains the scaling constant $b$.
This constant was calibrated for typical SDSS frames with objects
of average brightness. When determining the magnitudes of the
central peaks of the stacks, it would be appropriate to use the same
magnitude system as SDSS. In our case, however, the value of $b$
had to be recalibrated to account for extremely faint objects, which
would mean the recalibration of the whole SDSS magnitude system
as well. For this reason, we decided to use the conventional Pogson
magnitudes instead.

Because the co-added Stripe 82 images are sky-substracted, calibrated
for atmospheric extinction and remapped to a common zeropoint,
the transformation of counts into magnitudes is straightforward:
\begin{equation}
	\text{mag}_\lambda = -2.5 \log_{10}\text{counts} + 30 - A_\lambda 
\end{equation}
where $A_\lambda$ is the correction for the galactic extinction, discussed later in Section~\ref{sec:reddening}. 

We measure the brightness of the central peaks in a circular aperture with a radius of $5$ arcsec. The size of the aperture was chosen to be slightly larger than the double of FWHM of the central peaks ($1.5$ -- $2$ arcsec, depending on the band). To characterize the photometric uncertainties, we use the standard deviations provided by the jackknife stacks.

\subsection{Estimating the galactic extinction}
\label{sec:reddening}

We correct for the galactic extinction after stacking. We could have
applied an intensity rescale to each cut-out image prior stacking,
but that would have altered the noise characteristics of the images.
Instead, we correct for galactic extinction using an average value
of $ E(B-V) $. The average extinction is calculated from values of
the extinction in the directions of the DCS source coordinates. The
values of $ E(B-V) $ are taken from \citet{Schlegel 1998}. 
Table~\ref{tab:ext} summarizes the value of the galactic extinction
and the correction magnitudes for every imaging filter. The average
value of $ E(B-V) $ is equal for all three subsamples to two decimals.
Note, however, that the scatter in extinction values is almost as large
as the average extinction itself. Still, this only introduces a small
error in the final colour index estimates, as we show below.

\begin{table}
 \centering
  \begin{tabular}{ c c }
  \hline
   $ \left< E(B-V) \right> $ & $ \sigma_{E(B-V)} $ \\
 \hline
 \hline
 0.048 & 0.025\\

\end{tabular}
  \begin{tabular}{ c c c c c }
  \hline
   $ A_u $ & $ A_g $ & $ A_r $ & $ A_i $ & $ A_z $ \\
 \hline
 \hline
  0.25 & 0.18 & 0.13 & 0.10 & 0.07\\
 \hline
\end{tabular} 
  \caption{Average and standard deviation of the $ E(B-V) $ values of 
the DCS sample, with the correction magnitudes.}
 \label{tab:ext}
\end{table}

\subsection{Optical magnitudes}

The optical magnitudes of the central peaks are summarized in 
Table~\ref{tab:mags} for every SDSS filter and DCS subsample. The quoted 
magnitudes are already corrected for average foreground extinction, 
as explained in Section~\ref{sec:reddening}.

\begin{table}
 \centering
  \begin{tabular}{c r@{.}l @{ $\pm$ } r@{.}l r@{.}l @{ $\pm$ } r@{.}l r@{.}l @{ $\pm$ } r@{.}l}
  \hline
   Filter & \multicolumn{12}{c}{$S_{\text{int},1400}$ bin [mJy]} \\
	      & \multicolumn{4}{c}{$1-2$} & \multicolumn{4}{c}{$2-4$} & \multicolumn{4}{c}{$>4$} \\
 \hline
 \hline
 $u$ & 27&21& 2&63$^{*}$ & 26&20& 0&55 & 25&38& 0&38 \\
 $g$ & 25&57& 0&16 & 25&31& 0&13 & 25&18& 0&12 \\
 $r$ & 24&91& 0&19 & 24&71& 0&12 & 24&47& 0&09 \\
 $i$ & 24&29& 0&13 & 24&15& 0&12 & 23&82& 0&09 \\
 $z$ & 23&50& 0&15 & 23&54& 0&16 & 23&23& 0&12 \\
 
 \hline
\end{tabular}
  \caption{Optical magnitudes of the central peaks identified in the 
stacked images for every DCS subsample. The photometric calibration 
is described in Section~\ref{sec:photometry}. Values are corrected for 
average foreground galactic extinction. * Aperture radius: $ 3 $ arcsec.} 
\label{tab:mags}
\end{table}
 
According to these results, the average optical magnitudes of the
central peaks are about $ 1 $ mag fainter than the practical detection
limit of the Stripe 82 co-added catalogue. There is a slight increase
in optical brightness in all optical wavelength bands towards
higher radio fluxes. The error of the photometry strongly depends
on the imaging filter; the noise becomes really significant in the
ultraviolet.

\subsection{Radial profiles}
\label{sec:radprof}

\begin{figure*}
  \resizebox{17.4cm}{!}{\includegraphics{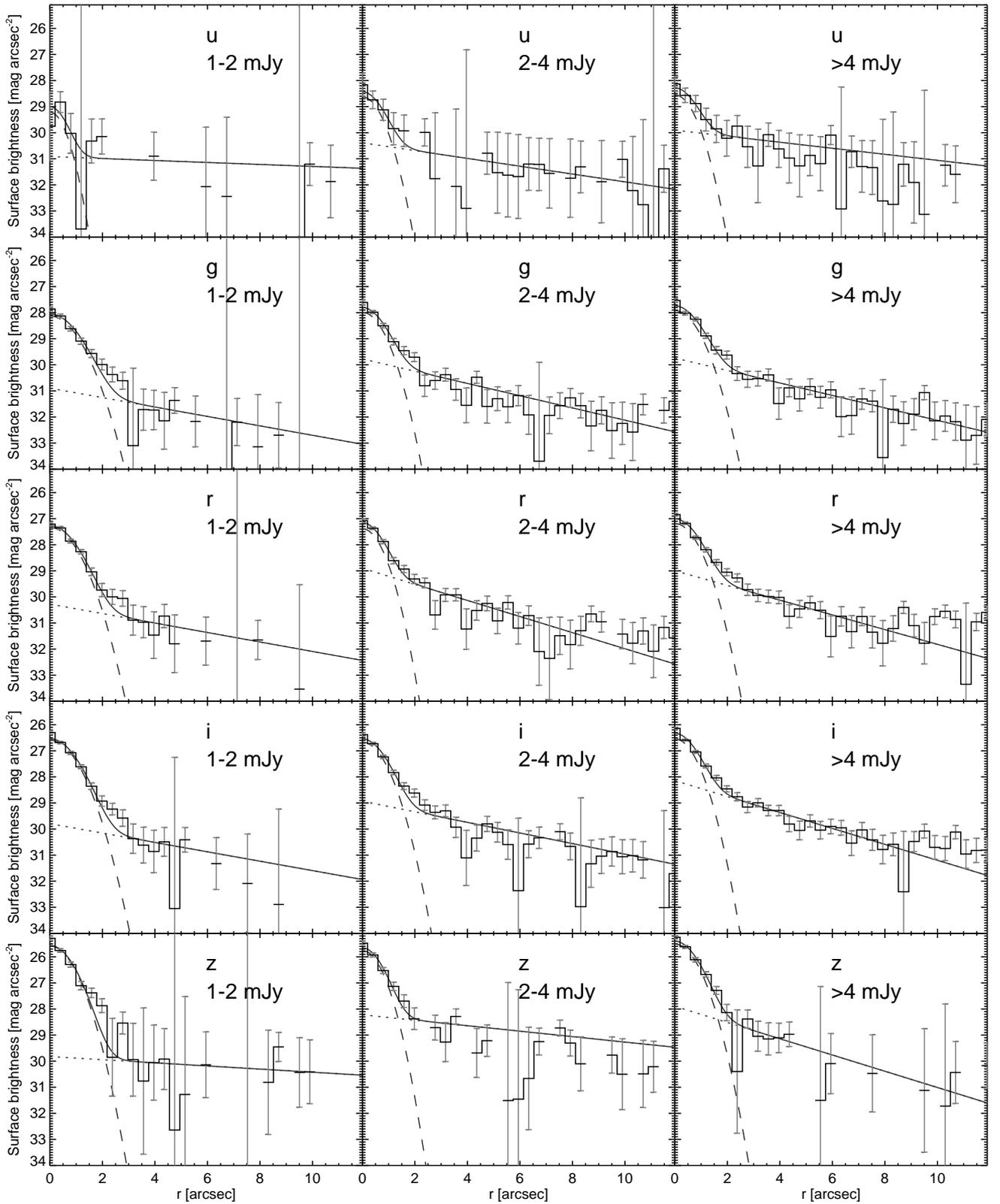}}
  \caption{Surface brightness profiles of the stacked images around 
the central peaks. The graphs in each column show the profiles for a given 
radio flux bin for all five SDSS imaging filters. Solid lines indicate the 
model fits; dashed lines are the Gaussian components; dotted lines are the 
exponential components. Error bars represent the variance of the values 
in each annulus.}  
  \label{fig:rad}
\end{figure*}

To determine the radial profiles of the stacked objects, we first determine the exact centres of the central peaks in the images by fitting two-dimensional Gaussians to the pixel values. Then we average the pixels in annuli of {$ \text{d}r = 0.4 $ arcsec} centred on the fitted positions of the peaks.

In Fig.~\ref{fig:rad} we plot the radial profiles of the background corrected stacks for every filter and DCS subsample. The central peaks of the profiles are clearly visible, just like the faint, extended components that decrease with radius.

We fit the profiles with the combination of a Gaussian and an exponential function. These fits are also displayed in Fig.~\ref{fig:rad} (solid lines). The fitting function is formulated the following way:
\begin{equation}
	I(r) = \frac{A}{\sqrt{2 \pi} \, \sigma} \exp{\left( -\frac{r^2}{2 \sigma^2} \right)} + B \, \exp{\left( - \frac{r}{r_0} \right)}
\end{equation}

The fitting is done in two phases. First, we fit the outer part
($ 3 < r < 8 $ arcsec) of the profiles with the exponential term to obtain the
value of $B$ and the scalelength $r_0$. Then, the resulting exponential
fit is subtracted from the original profile. Finally, the residuals are
fitted with Gaussians in the $0$--$5$ arcsec radius range. The fitted
parameters are summarized in Table~\ref{tab:radialfit}. The central PSF-like peaks
cut off around $ r \simeq 2 $ -- $ 3 $ arcsec, but the exponential component is
traceable to ca. $ r \simeq 15 $ arcsec. The $u$ profiles in the $1$--$2$ mJy and $ > 4 $ mJy
bins do not have good enough data quality to really trust the fitted values.

\begin{table}
 \centering
  \footnotesize{\begin{tabular}{c c   r@{.}l @{ $\pm$ } r@{.}l    r@{.}l @{ $\pm$ } r@{.}l    r@{.}l @{ $\pm$ } r@{.}l}
  \hline
       & & \multicolumn{12}{c}{$S_{int,1400}$ [mJy]} \\
	      & & \multicolumn{4}{c}{$1-2$} & \multicolumn{4}{c}{$2-4$} & \multicolumn{4}{c}{$>4$} \\
 \hline
 \hline
 
     \multirow{4}{*}{$u$}      & $A$ &   9&56 &   5&90 &  19&96 &   4&92 &  21&62 &   3&20 \\ 
      & $\sigma$ &   0&51 &   0&32 &   0&62 &   0&16 &   0&63 &   0&10 \\ 
      & $B$ &   1&57 &   0&82 &   2&55 &   1&59 &   3&97 &   2&23 \\ 
      & $r_0$ &  28&70 &  56&13 &   7&24 &   5&41 &   9&38 &   9&09 \\ 
     \hline
     \multirow{4}{*}{$g$}      & $A$ &  44&23 &   2&47 &  39&59 &   2&79 &  46&26 &   2&82 \\ 
      & $\sigma$ &   0&87 &   0&05 &   0&68 &   0&05 &   0&75 &   0&05 \\ 
      & $B$ &   1&59 &   0&90 &   4&47 &   1&46 &   4&61 &   1&26 \\ 
      & $r_0$ &   6&01 &   4&14 &   4&63 &   1&36 &   4&54 &   1&11 \\ 
     \hline
     \multirow{4}{*}{$r$}      & $A$ &  87&24 &   2&78 &  63&10 &   3&84 &  87&39 &   4&56 \\ 
      & $\sigma$ &   0&82 &   0&03 &   0&62 &   0&04 &   0&72 &   0&04 \\ 
      & $B$ &   2&82 &   1&04 &  10&00 &   3&41 &   9&16 &   2&05 \\ 
      & $r_0$ &   5&98 &   2&82 &   3&52 &   0&89 &   3&85 &   0&67 \\ 
     \hline
     \multirow{4}{*}{$i$}      & $A$ & 173&73 &   6&60 & 139&96 &   7&76 & 147&14 &   8&10 \\ 
      & $\sigma$ &   0&82 &   0&03 &   0&72 &   0&04 &   0&67 &   0&04 \\ 
      & $B$ &   4&39 &   2&13 &   9&71 &   3&04 &  20&09 &   3&47 \\ 
      & $r_0$ &   6&03 &   3&94 &   5&34 &   1&68 &   3&55 &   0&44 \\ 
     \hline
     \multirow{4}{*}{$z$}      & $A$ & 394&57 &  26&77 & 279&43 &  27&24 & 405&69 &  20&20 \\ 
      & $\sigma$ &   0&74 &   0&05 &   0&62 &   0&06 &   0&71 &   0&04 \\ 
      & $B$ &   4&27 &   2&13 &  18&69 &   8&40 &  25&27 &  11&34 \\ 
      & $r_0$ &  18&12 &  36&27 &  10&37 &   8&10 &   3&48 &   1&31 \\ 
     \hline
 \hline
\end{tabular}}
 \caption{Fitted parameters of the radial profiles: $A$ [nJy~arcsec$^{-1}$], 
 $\sigma$ [arcsec], $B$ [nJy arcsec$^{-2}$], $r_0$ [arcsec].}
\label{tab:radialfit}
\end{table}

\subsection{Possible origins of the exponential component}

The existence of an exponentially decreasing component of the radial profiles is obvious from Fig.~\ref{fig:rad}. The value of the scale parameter $r_0$ are listed in Table~\ref{tab:radialfit}. Excluding the low quality fits of the $u$ filter, the value of $r_0$ is consistently around $r_0 \simeq 3.5$ -- $5.5$~arcsec. 

It is very important to consider that the PSF of SDSS images already shows an extended component. To investigate the contribution of these extended wings to the radial profiles of our stacks, we applied our stacking algorithm to images of faint stars. The stacks of stars showed a similar exponential tail with a scale parameter $r_0$ in the range of $ 2 $ -- $ 3 $ arcsec, depending on the photometric band. These values are significantly smaller than the scale lengths derived from the stacks of our radio-selected objects. The radial profiles of our stacks clearly show an excess over the stellar PSF at the tails. For more discussion on the effect of the PSF wings on the radial profiles of stacked objects, see \citet{Zibetti 2004} and \citet{de Jong 2008}.

It is very risky to draw any conclusions from the observation of
the exponential component, especially in the absence of information
about the redshift of the radio-selected objects. Because the redshifts
are unknown, we cannot rescale our sources to the same physical
scale before the stacking. The physical scales could therefore be
very different between the objects that are stacked together.

However, it is interesting to see what the measured angular scalelength
translates to at different redshifts. If we make the assumption
that the majority of the DCS sample lies around $z \simeq 0.1$, the corresponding
scalelength is $6$-$10$~kpc. At $z \simeq 1$, the scalelength would
be $28$-$45$~kpc. The former range is consistent with the size of individual
galaxies, which might suggest that we see the host galaxies
of the quasars. The latter scalelength could correspond to the diameter
of merging systems. It can also be imagined that the excess
in surface brightness at large radii comes from undetected galaxies
clustered around the radio sources.

\subsection{Spectral energy distributions}
\label{sec:sed}

\begin{figure}
\includegraphics{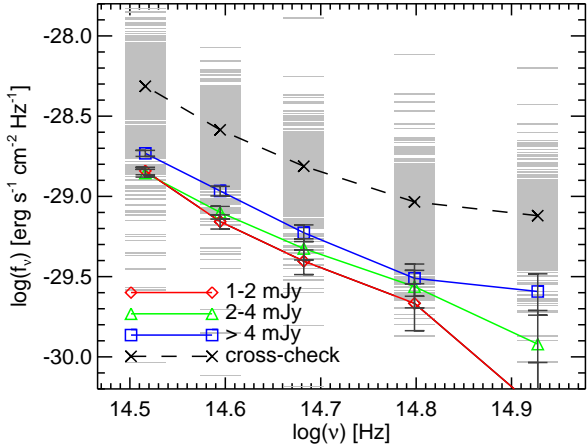}
  \caption{Optical SEDs (symbols connected with solid lines) of the central 
peaks identified in the stacked images for every DCS subsample. The short 
grey lines indicate the fluxes of the individual objects of the cross-check 
sample described in Section~\ref{sec:discussion}. The dashed line is the average SED 
of the cross-check sample.}
\label{fig:sed}
\end{figure}

One important finding is that the stacked objects have unusually red
average colours, as listed in Table~\ref{tab:colours}. The colour indices seem to be
independent from the apparent radio luminosity. In Fig.~\ref{fig:sed}, we plot
the SEDs of the central peaks of the stacked images for all three
DCS subsamples (solid lines with symbols). The optical spectral
indices ($\alpha_\nu$) are determined by fitting power-law functions to the
fluxes, and the results are listed in Table~\ref{tab:spidx}.

Because the redshift distribution of the DCS sample is unknown,
the colours of the stacked object might not correctly represent the
average spectral slope of the sample. In order to obtain the correct
colours, one should perform k-correction for each DCS source
individually -- obviously, this cannot be done in this study. However,
if we assume that the majority of the sources have a power-law SED
in the optical and ultraviolet regime, with similar spectral slopes, the
k-correction magnitudes in this case are independent of wavelength,
and hence the colour indices remain unchanged.

However, we note that if the objects had a redshift distribution 
similar to that of the LBDS sample mentioned in Section~\ref{sec:radiopoint} 
\citep{Waddington 2001}, that is, the majority of the objects had $ z < 1.5 $, 
based on the average spectral index of $\alpha_\nu = -2.5$, then the average 
k-correction would be $\sim 0.9$~mag in every band.

\begin{table}
 \centering
  \begin{tabular}{c r@{.}l @{ $\pm$ } r@{.}l r@{.}l @{ $\pm$ } r@{.}l r@{.}l @{ $\pm$ } r@{.}l}
  \hline
   Filter & \multicolumn{12}{c}{$S_{\text{int},1400}$ [mJy]} \\
	      & \multicolumn{4}{c}{$1-2$} & \multicolumn{4}{c}{$2-4$} & \multicolumn{4}{c}{$>4$} \\
 \hline
 \hline
 $g-r$ & 0&7&0&2 & 0&6&0&2 & 0&7&0&1 \\
 $r-i$ & 0&6&0&2 & 0&6&0&2 & 0&6&0&1 \\
 $i-z$ & 0&8&0&2 & 0&6&0&2 & 0&6&0&1 \\ 
\hline
\end{tabular}
  \caption{Optical colour indices of the central peaks 
  identified in the stacked images for every DCS subsample.}
\label{tab:colours}
\end{table}

\begin{table}
 \centering
  \begin{tabular}{lc}
  \hline
   $S_{\text{int},1400}$ & $\alpha_\nu$ \\
   bin [mJy] & \\
 \hline
 $1-2$ & $-2.9\pm 0.3$ \\
 $2-4$ & $-2.5\pm 0.1$ \\
 $>4$ & $-2.2\pm 0.3$ \\
\hline
\end{tabular}
  \caption{Average spectral indices of the central peaks.}
\label{tab:spidx}
\end{table}

\citet{Vanden Berk 2001} calculated composite spectra of more 
than 2200 SDSS quasars and they found $\alpha_\nu = -0.44$ for the spectral 
continuum. \citet{Ivezic 2002} determined spectral indices individually 
for 6868 quasar spectra with a mean of $\alpha_\nu = -0.45$. The 
bulk of their distribution is in the range of $-1 < \alpha_\nu < 0.2$. 
\citet{Gregg 2002} suggested $\alpha_\nu < -1$ as a definition of a red quasar. 
They found two sources with $\alpha_\nu \approx -3.7 $ and $\alpha_\nu \lesssim -4.6  $, which 
they called extraordinarily red. \citet{Richards 2006} defined Type~1 quasars by 
$\alpha = -0.5 \pm 0.3$ and reddened Type~1 quasars by $\alpha_\nu < -1$, 
which both have broad emission lines. The values for our 
stacked sources are in the range $ -2.9 \leq \alpha_\nu \leq -2.2 $. However, note 
that we cannot account for the effect of emission lines when calculating 
the spectral indices, and therefore we have $\alpha_\nu$ values from 
fitting of the entire spectrum (continuum and lines). Measurements of 
$\alpha$ by others usually based on the spectral continuum only.

If we account the red colours to intrinsic extinction, the values 
of $ -2.9 \leq \alpha_\nu \leq -2.2 $ require unusually high column densities of the 
obscuring dust. The absolute value of the spectral index becomes 
slightly smaller with increasing radio flux. This might suggest -- if 
we assume that the redshift distribution of the objects is narrow, or 
at least similar for all radio flux bins -- that the brighter radio sources 
are slightly less reddened on average.

\section{Discussion}
\label{sec:discussion}

Because the individual objects of the DCS sample are undetectable
in the optical images, we can only discuss the average optical properties
of the objects based on their stacked images. We hope that
these average values represent the original distribution of the sources
well. With the lack of direct detections, the homogeneity of the sample
cannot be proven. However, it is still worth comparing the DCS
sample with samples of very similar selection criteria, except that
the radio-selected objects are also detected in the optical.

\subsection{Construction of the cross-check sample}
\label{sec:cross-check}

In order to compare the average optical properties of the DCS objects
to better-known objects, we construct a cross-check sample of radio selected
sources, but in this case with identified optical counterparts.
We select FIRST radio point sources that have optical counterparts
within $1.5$~arcsec in the Stripe 82 catalogue, but not in the singly
observed (thus less faint) SDSS Data Release 6 (DR6) catalogue
within $3$~arcsec. The 1.5-arcsec matching radius was chosen in
accordance with \citet{Ivezic 2002}, who matched FIRST with
SDSS using the same separation limit. Based on their study, we
expect similar completeness ($\sim 85$ per cent), and a slightly higher
level of contamination than their $3$ per cent, because the larger
source density in the Stripe 82 co-added catalogue increases the
false-positive match rate. All other selection criteria for the FIRST
sources are the same as described in Section~\ref{sec:data}.

\begin{figure*}
  \includegraphics{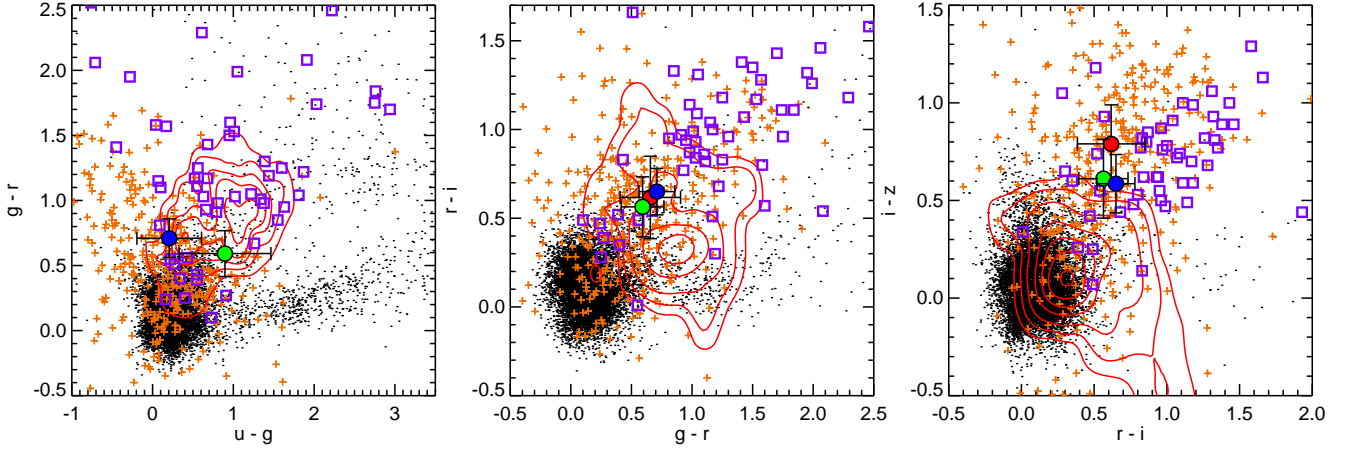}
  \caption{Colour-colour diagrams of the stacked objects (filled circles -- red: 
$1$--$2$~mJy, green: $2$--$4$~mJy, blue: $>4$~mJy), the cross-check sample 
(orange crosses), normal Type~1 quasars (black dots), red Type~1s (purple squares),
and Type~2s (red contour lines). See the description of the samples in 
Section~\ref{sec:compare}. A colour version of this plot is available on-line.}
  \label{fig:colour}
\end{figure*}

Because the co-added Stripe 82 is a significantly deeper survey
than the singly observed SDSS DR6, we expect to find objects that
are too faint to be detected in DR6 but are visible in the co-added
Stripe 82 images. Indeed, $1349$ FIRST sources were found with
optical magnitudes in the range $24 \geq m_i \geq 22$ mag. We apply
further restrictions on the photometric quality of the cross-check
sample (i.e. the error in $u$ and $g$ model magnitudes must be less
than $0.8$ and $0.5$~mag, respectively). The final sample consists of
$394$ co-added Stripe 82 sources, which are optical counterparts of
isolated, compact FIRST radio sources with integrated luminosities
larger than $S_{\text{int},1400} \geq 1 $~mJy.

We have published a catalogue on the multiwavelength photometric
properties of the cross-check objects, compiled from the
Stripe 82 co-added survey (optical), the FIRST survey (radio)
and two infrared surveys: the United Kingdom Infrared Telescope
(UKIRT) Infrared Deep Sky Survey (UKIDSS) and the Wide-field
Infrared Survey Explorer (WISE). The infrared cross-identification
is described in Section~\ref{sec:IR}. The catalogue is available on-line at
\url{http://www.vo.elte.hu/doublestacking}.

\begin{figure*}
\centering
\includegraphics{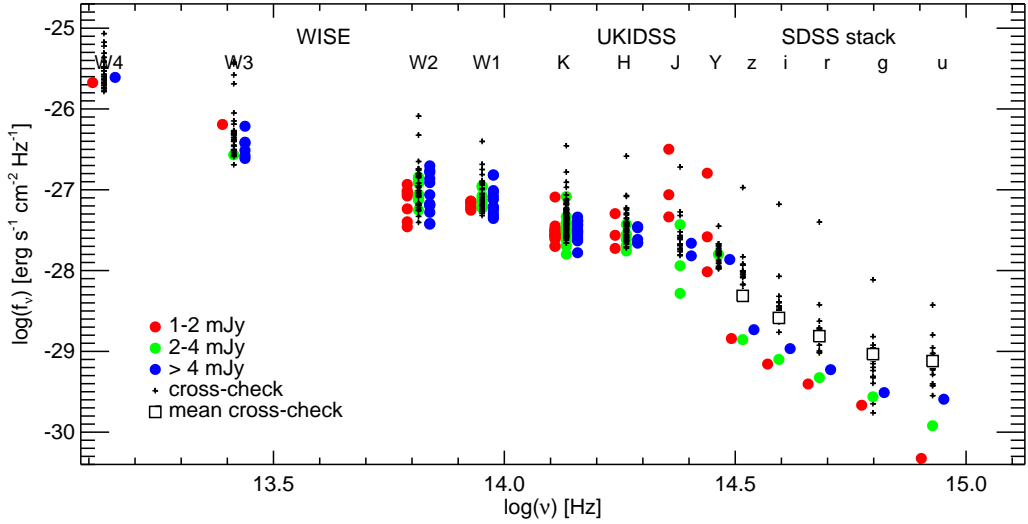}
  \caption{Optical and infrared SEDs of the DCS sample 
(coloured dots: red: $ 1 $ -- $ 2 $ mJy, green: $ 2 $ -- $ 4 $ mJy,
blue: $ >4 $ mJy subsamples) and the cross-check sample (black crosses). 
The data points in all infrared bands are individual detections, while 
the optical fluxes of the stacked objects are average values. We plot 
only those objects of the cross-check sample that have matching detections 
in the $K$ band. The average optical fluxes of the entire cross-check 
sample are plotted with black squares. (A colour version of this figure 
is available on-line.) }
\label{fig:sed_wide}
\end{figure*}

We plot the optical fluxes of the cross-check sample in Fig.~\ref{fig:sed} as grey dashes. The average fluxes of the cross-check sample (crosses connected with a dashed line) and the fluxes of the stacked objects (points connected with solid lines) are also plotted in the same plot. The average magnitudes calculated from the stacked images are about $1.2$--$1.6$ mag fainter than those of the cross-check sample, but the slopes of the SEDs are similar.

\subsection{Comparison with the cross-check sample and optically detected quasars}
\label{sec:compare}

To evaluate the average optical properties of the DCS objects we compare them to three samples of quasars. The first sample is the fifth edition of the SDSS quasar catalogue containing 105783 objects \citep{Schneider 2010}. The catalogue is composed of SDSS quasars with luminosities higher than $M_i < -22.0$, having at least one emission line with FWHM higher than $1000 ~ {\text{km}}~{\text{s}}^{-1}$ or showing interesting/complex absorption features. The quasar redshifts range from $0.065$ to $5.46$, with a median value of $1.49$. The majority of the sample can be classified as normal (unreddened) Type 1 quasars. The second quasar sample consists of 51 red Type 1 quasars \citep{Urrutia 2009}. \citeauthor{Urrutia 2009} matched the FIRST-2MASS red quasar survey \citep{Glikman 2007} with SDSS, and compiled this sample of spectroscopically confirmed red Type 1 quasars. The third sample contains 887 spectroscopically confirmed Type 2 quasars from SDSS with redshifts $z < 0.83$ \citep{Reyes 2008}. The stacked objects are about $ 1.4 $ mag fainter in the $i$ band than the average of the cross-check sample and the latter is $ 3 $ mag fainter than the red Type 1 sample.

In Fig.~\ref{fig:colour} we plot the optical colour-colour diagrams of the stacked objects, the cross-check sample and the three quasar samples. It is clear that \textit{on average} our DCS sample is significantly redder than normal Type~1 quasars. The $r-i$ and $i-z$ colours suggest that the DCS sample is more likely to consist of reddened Type~1s than Type~2s.

\subsection{Infrared counterparts}
\label{sec:IR}

If we accept the assumption that our radio sources are AGNs heavily
obscured by dust surrounding the central regions, it is straightforward
to look for excess in their infrared emission. For this reason,
we looked for infrared counterparts to the DCS sample in the extragalactic
near-infrared UKIDSS Large Area Survey (LAS) DR7
catalogue \citep{Dye 2006} and in the WISE preliminary release
mid-infrared source catalogue \citep{Wright 2010}. Within the sky
regions overlapping with SDSS Stripe 82, about $4$ per cent of the
radio sources ($72$ objects) were detected by the UKIDDS in any of
the infrared bands, using a matching radius of $1.5$~arcsec. Only a few
dozen objects have matching counterparts in the WISE catalogue
in at least one band. This detection rate is about $5$ per cent. The
infrared magnitudes are close to the detection limits of both surveys
and the scatter of the magnitudes is small. This means that only the
very bright end of the sample is detected in infrared.

The cross-check sample has a much higher detection rate, with
$123$ sources detected in the near-infrared ($37$ per cent) and $33$ in the
mid-infrared ($32$ per cent). The distribution of the infrared magnitudes
is similar to that of the DCS sample, with cut-off at the faint
end because of the magnitude limit. In Fig.~\ref{fig:sed_wide}, we plot the optical to
mid-infrared SED of our samples: the individually detected objects
of the DCS sample (colour dots) and the cross-check sample (black
crosses). The data points in all infrared bands are individual detections,
while the optical fluxes of the stacked objects are average
values for the three DCS subsamples. We plot the average optical
fluxes of the cross-check sample with black squares.

The cross-check sample has an average SED with a slope of
$ \alpha_\nu \simeq -2.0 $ (see Fig.~\ref{fig:sed}), less steep than the slope of the average
optical SED of the DCS sample. Exact physical interpretation is
difficult, however, as we do not have information on the composition
of the samples. If we assume that the objects in the samples are
intrinsically similar, the gradual steepening of the spectral slopes
with decreasing optical brightness can be attributed to obscuration
by dust.

\begin{figure}
\includegraphics{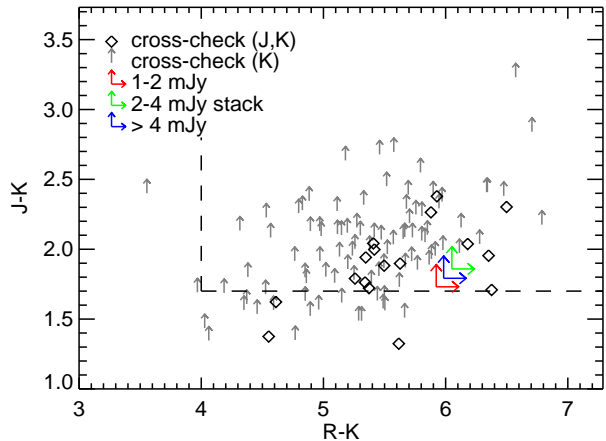}
  \caption{Colour indices of the cross-check sample and the DCS sample 
in the $R-K$;$J-K$ plane. Diamonds denote cross-check sample objects with 
valid $J$ band data, arrows denote cross-check objects with the $J$ band 
lower limit only and coloured arrows denote stacked objects (red: $ 1 $ 
-- $ 2 $ mJy, green: $ 2 $ -- $ 4 $ mJy, blue: $ >4 $ mJy subsamples). 
The arrows mark the lower limits on $R-K$ and $J-K$. The dashed lines bound 
the upper right region, where $50$ per cent of the sources are red quasars, 
according to \citet{Glikman 2004}.}
\label{fig:RJK}
\end{figure}

\citet{Glikman 2004} constructed a sample of bright near-infrared
sources that are detected at radio wavelengths but unidentified on 
the Palomar Observatory Sky Survey (POSS) plates, with the aim 
of finding dust-obscured quasars. They have defined a region in 
the $R-K$, $J-K$ colour plane, in which $50$ per cent of the 
radio-selected objects are highly reddened quasars ($J-K > 1.7$ and $R-K > 4$). 
To compare our samples to these criteria, we converted 
SDSS magnitudes to Johnson $R$ using the equations of \citet{Jester 2005}. 
There are $17$ sources of our cross-check sample that have 
detections in both $J$ and $K$ infrared bands. For objects not detected 
in the $J$ band (the majority), we assume a lower limit of the $J-K$ 
colour index using the limiting magnitude of the survey. In Fig. \ref{fig:RJK}, 
we plot the cross-check objects in the $R-K$, $J-K$ colour plane. 
Only a few objects have valid $J$ magnitudes (diamonds); object 
with lower limits in $J$ are plotted with grey arrows. The majority 
of our sources ($77$ per cent) are in the red quasar region of the plane.

Only a few objects of the DCS sample were detected in infrared
individually. It is very likely that the optical flux distribution of these
objects differs significantly from the distribution of the entire DCS
sample. Because the DCS sources are undetected in the optical, we
can easily determine an upper limit on their $R$ fluxes; they should
be fainter than the Stripe 82 co-added survey limit ($R \gtrsim 24.2$).

Therefore, we are able to estimate a lower limit on the
$R-K$ colour of the DCS sources with infrared detections. We
also calculate a lower limit on $J-K$ in the absence of reliable $J$ band
fluxes. The results are plotted in Fig.~\ref{fig:RJK} with coloured symbols
(arrows), one for each DCS subsample. It is apparent that the $R-K$
values are in the red quasar region, even if we assume brighter $R$
fluxes. The $J-K$ colours are also in that region, but much closer
to the border.

It is hard to estimate the possible fraction of red quasars in our
samples based on the available infrared data. As for the cross-check
sample, with a $34$ per cent detection rate and with at least $77$ per
cent of the infrared-matched objects lying in the red quasar region
of the $R-K$, $J-K$ plane, we find that at least $13$ per cent of all
objects are undoubtedly dust-obscured quasars. However, this is a
weak lower limit on their number, because we only see the brightest
infrared counterparts as a result of the relatively high infrared flux
limit of the observations.

\section{Summary}

We have presented an analysis based on an image stacking technique,
to reveal the visible wavelength light from the unresolved
sources of the FIRST radio survey that remain undetected in the
SDSS Stripe 82 co-added catalogue. The sample consists of \samplesize
objects, which have been divided into three subsamples according
to the radio flux. We stacked cut-out images centred on the object
coordinates from the SDSS Stripe 82 co-added survey. The main
results of our study are as follows.

\begin{enumerate}
  \item We have presented an image stacking algorithm with an efficient 
masking technique to reveal the average optical emission of 
\samplesize optically undetected sources (DCS sample) selected based on 
the VLA FIRST radio survey.
  \item We have detected the average optical emission of these radio 
sources (with $26.6$ mag $5\sigma$ detection limit in $r$ band), and we
have calculated the $ugriz$ magnitudes of the peaks apparent in the 
centres of the stacked images (see Table~\ref{tab:mags}). The colours of these 
peaks imply a steep, red optical spectrum with spectral indices 
of $ -2.9 \leq \alpha_\nu \leq -2.2 $.
  \item The average surface brightness profiles of the stacked objects 
show Gaussian peaks in the centres, which continue in outwards 
fading exponential components traceable to $ \sim 15 $ arcsec. The 
surface brightness of these components peaks at approximately 
$2$--$3$~mag~arcsec$^{-2}$ fainter than the peak surface brightness of the 
central point-like components.
  \item We have created a radio-selected sample with faint optical 
detections in the SDSS Stripe 82 catalogue (\textit{cross-check sample}), 
similar to the DCS sample. We have identified infrared counterparts 
to the cross-check objects from the UKIDSS near-infrared 
and WISE mid-infrared source catalogue (with $37$ per cent detection 
rate), and we have composed a catalogue of the optical, infrared 
and radio properties of the sample.
  \item We have compared the optical colour indices of the radio-selected 
and cross-check objects with various spectroscopically 
identified quasar samples. We have concluded that the distribution 
of the colours are similar to that of dust-reddened Type~1 quasars, 
although there is large scatter in the data.
  \item We have identified the infrared counterparts of the DCS 
sources with a $5$ per cent detection rate. We have investigated the 
distribution of the cross-check and DCS sources in the $R-K$, $J-K$ 
colour plane, and we have found that the majority of the 
objects lie in a region where $\sim 50$ per cent of the sources are red quasars.
  \item Consequently, the sources of the DCS sample and the cross-check 
sample are very good red quasar candidates, suitable for future 
deeper observations and quasar studies. We assume that the 
forthcoming optical and infrared survey telescopes, such as the
Large Synoptic Survey Telescope (LSST) and the James Webb
Space Telescope (JWST), might discover many more obscured
quasars than previously expected.
\end{enumerate}
  
\section*{Acknowledgements}

The authors would like to thank S\'{a}ndor Frey for his help in the field of radio astronomy.

This work was supported by the following Hungarian grants: NKTH: Pol\'anyi, OTKA-80177, OTKA-103244 and KCKHA005.

The Project is supported by the European Union and co-financed by the European Social Fund (grant agreement no. T\'AMOP 4.2.1./B-09/1/KMR-2010-0003).

Funding for the SDSS and SDSS-II has been provided by the Alfred P. Sloan Foundation, the Participating Institutions, the National Science Foundation, the U.S. Department of Energy, the National Aeronautics and Space Administration, the Japanese Monbukagakusho, the Max Planck Society, and the Higher Education Funding Council for England. The SDSS Web Site is http://www.sdss.org/.

The SDSS is managed by the Astrophysical Research Consortium for the Participating Institutions. The Participating Institutions are the American Museum of Natural History, Astrophysical Institute Potsdam, University of Basel, University of Cambridge, CaseWestern Reserve University, University of Chicago, Drexel University, Fermilab, the Institute for Advanced Study, the Japan Participation Group, Johns Hopkins University, the Joint Institute for Nuclear Astrophysics, the Kavli Institute for Particle Astrophysics and Cosmology, the Korean Scientist Group, the Chinese Academy of Sciences (LAMOST), Los Alamos National Laboratory, the Max-Planck-Institute for Astronomy (MPIA), the Max-Planck-Institute for Astrophysics (MPA), New Mexico State University, Ohio State University, University of Pittsburgh, University of Portsmouth, Princeton University, the United States Naval Observatory, and the University of Washington.

This publication makes use of data products from the VLA FIRST survey, which is supported in part under the auspices of the Department of Energy by Lawrence Livermore National Laboratory under contract W-7405-ENG-48 and the Institute for Geophysics and Planetary Physics.

The UKIDSS project is defined in \citet{Lawrence 2007}. UKIDSS uses the UKIRT Wide Field Camera (WFCAM; \citet{Casali 2007}). The photometric system is described in \citet{Hewett 2006}, and the calibration is described in  \citet{Hodgkin 2009}. The pipeline processing and science archive are described in \citet{Hambly 2008}.

This publication makes use of data products from the Wide-field Infrared Survey Explorer, which is a joint project of the University of California, Los Angeles, and the Jet Propulsion Laboratory/California Institute of Technology, funded by the National Aeronautics and Space Administration.

\label{lastpage}

\end{document}